\documentclass[letterpaper,11pt]{article}

\usepackage[utf8]{inputenc}
\usepackage[T1]{fontenc}

\usepackage{jcapmod}

\usepackage{amsmath,amsfonts,amssymb}
\usepackage{mathtools,physics,bm}
\usepackage{textalpha}

\usepackage{hyperref}
\usepackage{graphicx}

\usepackage[letterpaper,margin=2.55cm]{geometry}

\numberwithin{equation}{section}

\def\sqrtb{\mathpalette\DHLhksqrt}
\def\DHLhksqrt#1#2{%
\setbox0=\hbox{$#1\sqrt{#2\,}$}\dimen0=\ht0
\advance\dimen0-0.2\ht0
\setbox2=\hbox{\vrule height\ht0 depth -\dimen0}%
{\box0\lower0.4pt\box2}}

\newcommand{\GN}{G_{\scriptscriptstyle\mathrm{N}}}
\newcommand{\lp}{\ell_{\scriptscriptstyle\mathrm{P}}}
\newcommand{\mpl}{m_{\scriptscriptstyle\mathrm{P}}}
\newcommand{\RS}{r_{\scriptscriptstyle\mathrm{S}}}
\makeatletter
\DeclareFontFamily{OMX}{MnSymbolE}{}
\DeclareSymbolFont{MnLargeSymbols}{OMX}{MnSymbolE}{m}{n}
\SetSymbolFont{MnLargeSymbols}{bold}{OMX}{MnSymbolE}{b}{n}
\DeclareFontShape{OMX}{MnSymbolE}{m}{n}{
    <-6>  MnSymbolE5
   <6-7>  MnSymbolE6
   <7-8>  MnSymbolE7
   <8-9>  MnSymbolE8
   <9-10> MnSymbolE9
  <10-12> MnSymbolE10
  <12->   MnSymbolE12
}{}
\DeclareFontShape{OMX}{MnSymbolE}{b}{n}{
    <-6>  MnSymbolE-Bold5
   <6-7>  MnSymbolE-Bold6
   <7-8>  MnSymbolE-Bold7
   <8-9>  MnSymbolE-Bold8
   <9-10> MnSymbolE-Bold9
  <10-12> MnSymbolE-Bold10
  <12->   MnSymbolE-Bold12
}{}

\let\llangle\@undefined
\let\rrangle\@undefined
\DeclareMathDelimiter{\llangle}{\mathopen}%
                     {MnLargeSymbols}{'164}{MnLargeSymbols}{'164}
\DeclareMathDelimiter{\rrangle}{\mathclose}%
                     {MnLargeSymbols}{'171}{MnLargeSymbols}{'171}
\makeatother

\begin{document}

%\pagenumbering{roman}

\begin{titlepage}

\baselineskip=15.5pt \thispagestyle{empty}

%%% TITLE %%%
\begin{center}
    %{\fontsize{18}{24}\selectfont \bfseries Complex Powerballs: \\\vspace*{.2cm}
    %{\it Conformal Cores of Quantum Black Holes in Quadratic Gravity}}
    {\fontsize{18}{24}\selectfont \bfseries Conformal Cores of Quantum Black Holes\\\vspace*{.2cm} in Quadratic Gravity}
\end{center}

\vspace{0.1cm}

%%% AUTHORS %%%
\begin{center}
    {\fontsize{12}{18}\selectfont Ruolin Liu,$^{1,2,3}$ Jerome Quintin,$^{4,2,3}$ and Niayesh Afshordi,$^{1,2,3}$}
\end{center}

%%% AFFILIATIONS %%%
\begin{center}
    \vskip8pt
    \textsl{$^1$ Department of Physics and Astronomy, University of Waterloo, Waterloo, ON N2L 3G1, Canada}\\
    \vskip4pt
    \textsl{$^2$ Perimeter Institute for Theoretical Physics, Waterloo, ON N2L 2Y5, Canada}\\
    \vskip4pt
    \textsl{$^3$ Waterloo Centre for Astrophysics, University of Waterloo, Waterloo, ON N2L 3G1, Canada}\\
    \vskip4pt
    \textsl{$^4$ Department of Applied Mathematics, University of Waterloo, Waterloo, ON N2L 3G1, Canada}
\end{center}

\vspace{1.2cm}

%%% ABSTRACT %%%
\hrule
\vspace{0.3cm}
\noindent {\bf Abstract}\\[0.1cm]
We explore the possibility that quadratic gravity, as a renormalizable theory, describes the interior of quantum black holes. We find new exact power-law solutions to pure quadratic gravity under spherical symmetry, which are complex valued. The resulting solutions, dubbed powerballs, are horizonless compact objects that become Schwarzschild-like a small distance (of the order of the Planck length) outside the would-be Schwarzschild horizon. We present a description of the global eternal geometry, whose right and left exteriors are Lorentzian and Euclidean Schwarzschild-like regions, respectively, while the complex interior is a form of spiraling spacetime. We compute the total on-shell action integral as a saddle point to a gravitational path integral and discuss the Lorentzian and Euclidean interpretations thereof.
\vskip10pt
\hrule
\vskip10pt

\end{titlepage}

%%% TABLE OF CONTENTS %%%
%% some of the options below can be commented out depending on preferred spacing and page numbering

%\thispagestyle{empty}
\tableofcontents
%\newpage
\pagenumbering{arabic}
\setcounter{page}{1}
%\clearpage

%\pagenumbering{arabic}
%\thispagestyle{empty}
%\setcounter{page}{1}
%\tableofcontents

%%%%%%%%%%%%%%%%%%%%%%%%%%%%%%%%%%%%%%%%%%%%%%%%%%%
%%%%%%%%%%%%%%%%%%%%%%%%%%%%%%%%%%%%%%%%%%%%%%%%%%%

\section{Introduction}

In general relativity (GR), spacetime singularities are ubiquitous (e.g., \cite{Penrose:1964wq}). The prototypical example is a Schwarzschild black hole, which has a scalar curvature singularity at its core. While GR can be treated as an effective field theory (EFT) at low curvature (e.g., \cite{Donoghue:1994dn,Donoghue:1995cz,Burgess:2003jk,Donoghue:2012zc,Donoghue:2017pgk,Ruhdorfer:2019qmk,Donoghue:2022eay}), pure GR is nonrenormalizable \cite{tHooft:1974toh,Christensen:1979iy,Goroff:1985th}, hence it becomes computationally intractable as one approaches a classical singularity, i.e., as the curvature blows up.

One thus expects GR to be modified in the ultraviolet (UV). In particular, a proper UV-complete theory of gravity should be either finite or renormalizable. Several approaches to this problem exist, such as string theory (e.g., \cite{Schwarz:1982jn}), loop quantum gravity (e.g., \cite{Ashtekar:1986yd,Rovelli:1994ge,Rovelli:1997yv,Ashtekar:2004eh}), causal sets (e.g., \cite{Bombelli:1987aa,Sorkin:2003bx,Dowker:2005tz,Surya:2019ndm}), causal dynamical triangulations (e.g., \cite{Ambjorn:2005db,Ambjorn:1998xu,Ambjorn:2012jv,Loll:2019rdj}), and asymptotic safety (e.g., \cite{Niedermaier:2006wt,Codello:2008vh,Benedetti:2009rx,Eichhorn:2018yfc,Donoghue:2019clr}). A more minimalistic and anterior theory is quadratic gravity \cite{Stelle:1976gc,Stelle:1977ry}, which consists of the standard Einstein-Hilbert action of GR with the addition of terms quadratic in curvature (see, e.g., \cite{Alvarez-Gaume:2015rwa,Salvio:2018crh} for reviews). The pioneering work of Stelle \cite{Stelle:1976gc} demonstrated that quadratic gravity is indeed renormalizable when treated perturbatively at the one-loop level. The renormalizability of quadratic gravity arises from the higher-derivative terms, which improve the UV behavior of the graviton propagators, rendering the divergences manageable within a quantum framework. Furthermore, the coupling constants of quadratic gravity as in any Wilsonian quantum field theories should run under renormalization group (RG) flow, and they appear to have a UV fixed point (see, e.g., \cite{Fradkin:1981iu,Avramidi:1985ki,Codello:2006in,Niedermaier:2009zz,Niedermaier:2010zz,Ohta:2013uca} and \cite{Holdom:2015kbf,Holdom:2016xfn,Buccio:2024hys,Buccio:2024omv,Kawai:2024aim} more recently). However, a massive ghost appears in this theory at the expense of renormalizability. Notably, the presence of fourth-order derivative terms in the equations of motion leads to the emergence of ghost states with negative norm or negative kinetic energy (see, e.g., \cite{Stelle:1977ry,Hinterbichler:2015soa}). These ghost states violate unitarity, posing a fundamental problem for the physical viability of the theory. Various approaches have been proposed to address this issue, such as treating the ghost states via indefinite metric quantization or exploring nonperturbative formulations (see, e.g., \cite{Hawking:2001yt,Mannheim:2004qz,Mannheim:2006rd,Bender:2007wu,Donoghue:2017fvm,Donoghue:2018izj,Donoghue:2019fcb,Donoghue:2019ecz,Donoghue:2020mdd,Donoghue:2021eto,Donoghue:2021meq,Donoghue:2021cza,Holdom:2023usn,Holdom:2024cfq,Holdom:2024onr}), but a universally accepted solution remains elusive. Nevertheless, quadratic gravity has been the subject of a lot of work in the context of (primordial) cosmology, strong gravity, and quantum gravity (e.g., \cite{Clunan:2009er,Deruelle:2010kf,Deruelle:2012xv,Anselmi:2020lpp,Anselmi:2021dag,Anselmi:2021rye,DeFelice:2023psw,Salvio:2017xul,Salvio:2017qkx,Salvio:2019ewf,Salvio:2024joi,Held:2023aap,Lehners:2019ibe,Lehners:2023fud,Bonanno:2024fcv,Buoninfante:2024oyi,Belokurov:2021oer,Belokurov:2024pjr,Edelstein:2021jyu,Edelstein:2024jzu}).

One may imagine that different approaches to quantum gravity might only manifest their differences at high curvature regimes, e.g., near singularities of classical GR. However, general arguments suggest that a quantum theory of gravity could couple the ultraviolet to the infrared physics nonperturbatively (e.g., \cite{Berglund:2022qcc}). The most famous example of this property is the {\it black hole information paradox} (e.g., \cite{Hawking:1976ra,Mathur:2009hf,Almheiri:2012rt}). It suggests quantum effects should become important on the scale of black hole event horizons, in order to enable unitary Hawking evaporation, even though the spacetime curvatures (at least in GR) remain small. This paradox motivates looking into solutions for macroscopic black holes in a UV-complete theory of quantum gravity (potentially testable in gravitational wave observations; see, e.g., \cite{Cardoso:2016rao,Cardoso:2016oxy,Abedi:2016hgu,Cardoso:2017cqb,Wang:2018gin,Abedi:2020ujo}).

Specifically in the context of black hole physics, quadratic gravity as a potential UV-complete theory leads to solutions that differ from the standard spherically symmetric solutions of GR, leading to new black hole solutions with distinct properties (see, e.g., \cite{Lu:2015cqa,Lu:2015psa,Lu:2015tle,Stelle:2017bdu,Lu:2017kzi}). There is also the possibility that quadratic gravity yields horizonless objects that are not quite black holes, such as 2-2 holes as proposed in \cite{Holdom:2016nek,Holdom:2022zzo} (see also \cite{Holdom:2019bdv,Aydemir:2020xfd,Holdom:2020onl,Holdom:2022npq,Holdom:2022fsm}).

In a similar vein to \cite{Borissova:2022jqj}, which explored scale-invariant cores of RG-inspired quantum black holes, this work shall make the postulate that the cores of `quantum black holes' should be conformally invariant. Specifically, we will explore the possibility that pure quadratic gravity describes high-curvature `black hole' interiors. The solution will be resolutely quantum mechanical and horizonless. The `outside' will resemble a Schwarzschild black hole, but at a Planck length above the would-be Schwarzschild horizon one would enter into a \emph{powerball}, a complex spacetime geometry that, as we will argue, makes sense from a path integral formulation of (quantum) gravity.

\paragraph*{Outline:} We start by deriving power-law solutions to pure quadratic gravity in Sec.~\ref{sec:powerSol}, thus defining complex powerballs. We explain in Sec.~\ref{sec:core} in what sense powerballs can be thought of as cores of (quantum) black holes. The geometry of the resulting spacetime is presented in Sec.~\ref{sec:geo}, and Sec.~\ref{sec:integral} is devoted to computing the action integral of every spacetime region. The gravitational path integral is discussed in Sec.~\ref{sec:path}, and the results are summarized and further discussed in Sec.~\ref{sec:concl}.

\paragraph*{Notation and conventions:} We set $c=\hbar=k_{\scriptscriptstyle\mathrm{B}}=1$ and shall equivalently use Newton's gravitational constant or the Planck length/mass following $\GN\equiv\lp^2\equiv\mpl^{-2}$. Greek tensorial indices run over $3+1$ spacetime coordinates and are raised/lowered with the metric tensor $g_{\mu\nu}$, regardless of the metric signature. Latin tensorial indices run over the coordinates of codimension-$1$ hypersurfaces and are raised/lowered with the induced metric tensor $h_{ab}$, once more regardless of the metric signature.

\section{Complex powerballs in pure quadratic gravity}\label{sec:powerSol}

The theory of quadratic gravity can be written in various ways due to the linear relations among quadratic invariants such as the Kretschmann scalar\footnote{A tensor `squared' is meant to be contracted with itself, e.g., $R_{\mu\nu\rho\sigma}^2=R_{\mu\nu\rho\sigma}R^{\mu\nu\rho\sigma}$ for the Riemann tensor.} $\mathcal{K}\equiv R_{\mu\nu\rho\sigma}^2$, the Ricci tensor squared $R_{\mu\nu}^2$ (where $R_{\mu\nu}=R^\rho{}_{\mu\rho\nu}$), the Ricci scalar squared $R^2$ (where $R=R^\mu{}_\mu$), the Weyl tensor squared $C_{\mu\nu\rho\sigma}^2=\mathcal{K}-2R_{\mu\nu}^2+R^2/3$, and the Gauss-Bonnet (topological) invariant $\mathcal{G}=\mathcal{K}-4R_{\mu\nu}^2+R^2$. One expression is
\begin{equation}
    S_\textrm{quad-grav}=\int\mathrm{d}^4x\,\sqrtb{-g}\left(\frac{1}{16\pi\GN}R-\Lambda+\frac{\omega}{3\sigma}R^2-\frac{1}{2\sigma}C_{\mu\nu\rho\sigma}^2\right)\,,
\end{equation}
where $\GN$ is Newton's gravitational constant, $\Lambda$ is a cosmological constant, and $\sigma$ and $\omega$ are dimensionless coupling constants that run under RG flow ($\GN$ and $\Lambda$ may likewise flow; see, e.g., \cite{Fradkin:1981iu,Avramidi:1985ki,Codello:2006in,Niedermaier:2009zz,Niedermaier:2010zz,Ohta:2013uca}). Henceforth, however, we will ignore the running of coupling constants as a first approximation (i.e., we just treat them as constants; we comment on this in the discussion section).
At high curvature or high momentum, the quadratic invariants should dominate the action of quadratic gravity (e.g., near the UV fixed point). A theory of pure quadratic gravity could thus be written as
\begin{equation}
    S_\textrm{pure-quad}=\int\mathrm{d}^4x\,\sqrtb{-g}\left(\frac{\omega}{3\sigma}R^2-\frac{1}{2\sigma}C_{\mu\nu\rho\sigma}^2\right)\,.\label{eq:Spurequad}
\end{equation}
The premise for this work is that such an action could accurately describe (quantum) gravity inside `quantum black holes'. In particular, if the spherically symmetric solution is Ricci flat (as we will shortly see), one has pure Weyl gravity and the theory is manifestly conformally invariant (see, e.g., \cite{Kazanas:1988qa,Mannheim:2011ds}).

Let us consider a spherically symmetric, stationary metric ansatz given by
\begin{equation}
    g_{\mu\nu}\dd x^\mu\dd x^\nu=-A(r)\dd t^2+B(r)\dd r^2+r^2\dd\Omega_{(2)}^2\,,\label{eq:gmunu}
\end{equation}
where $\dd\Omega_{(2)}^2=\dd\theta^2+\sin^2(\theta)\dd\phi^2$ is the line element on the unit $2$-sphere. Here, $A(r)$ and $B(r)$ could be arbitrary functions of the radial coordinate $r$, but we will be interested in exploring potential power-law solutions of the form
\begin{equation}
    A(r)=ar^\alpha\,,\qquad B(r)=br^\beta\,,\label{eq:power-law}
\end{equation}
where at this point $a,b,\alpha$, and $\beta$ are unspecified constants (which are not even necessarily real as we will shortly see).
For simplicity for the time being, we set $\GN=1$, but we will reinsert units later.
Upon varying the action of pure quadratic gravity \eqref{eq:Spurequad} with respect to $A$ and $B$, one obtains two equations of motion, which for the power-law ansatz \eqref{eq:power-law} reduce to
\begin{subequations}\label{eq:EOMs}
\begin{align}
    0=&-2 \omega  \left(\alpha ^2-(\alpha +4) \beta +2 \alpha +4\right) \left(\alpha ^2-\alpha  (\beta -2)-4 (\beta  (3
   \beta +7)+5)\right)\nonumber\\
   &+(\alpha -2) (\alpha -4 \beta -2) (\alpha -\beta -2) (\alpha +3 \beta -2)+16 b^2 (2 \omega
   -1) r^{2 \beta }-192 b \omega  r^{\beta }\\
   0=&-2 \omega  \left(\alpha ^2-(\alpha +4) \beta +2 \alpha +4\right) \left(\alpha ^2+3 (\alpha +4) \beta +2 \alpha
   +28\right)\nonumber\\
   &+(\alpha -2)^2 (\alpha -\beta -2) (\alpha +3 \beta -2)+16 b^2 (2 \omega -1) r^{2 \beta }+192 b
   \omega  r^{\beta }\,,
\end{align}
\end{subequations}
as long as $a,b,\sigma\neq 0$.
In an effort to solve these two equations, we notice that the $r^\beta$ terms must be eliminated to make $\alpha$ and $\beta$ independent of $r$. Thus, either the coefficients of each $r^\beta$ term and other remaining constant terms vanish, or the power of $r^\beta$ is zero. While there is no solution that can satisfy the former, the latter implies $\beta=0$, in which case the equations are readily solved if $\alpha=0$, $b=1$ or if
\begin{equation}
    \alpha=\alpha_\mp\equiv\frac{1}{2}\left(1\mp i \sqrtb{15}\right)\,,\qquad b=b_\mp\equiv\frac{3}{8}\left(1\mp i\sqrtb{15}\right)\,.\label{eq:solutions}
\end{equation}
The first solution ($\alpha=0$, $b=1$) represents the trivial Minkowski spacetime solution, but the other two (complex) solutions are nontrivial (they are complex conjugate of one another).
We will call such solutions \emph{complex powerballs}.
Notice that these two solutions are independent of the coupling constants $\sigma$ and $\omega$ of quadratic gravity, and they indicate that $g_{tt}=ar^{\alpha_\mp}$ has a complex power with an arbitrary (potentially complex) coefficient $a$, while $g_{rr}=b_\mp$ is a complex constant.
One can check that these complex solutions further have the property of being Ricci flat, i.e., $R=0$, while the Ricci tensor squared is complex,
\begin{equation}
    R_{\mu\nu}^2=-\frac{2\left(3\pm i\sqrtb{15}\right)}{9r^4}\,,
\end{equation}
and the Weyl tensor squared is real,
\begin{equation}
    C_{\mu\nu\rho\sigma}^2=\frac{16}{3r^4}\,.\label{eq:C2s}
\end{equation}
Note that this diverges as $r\searrow 0$, but it is softer than the $\propto r^{-6}$ singularity of Schwarzschild black holes.
It can be proved that this pair of complex solutions is consistent with a general expression for stationary black holes considering the Tolman-Oppenheimer-Volkoff equation under the Ricci-flat condition.

These complex metrics may similarly be found by demanding Ricci flatness in the context of pure quadratic gravity. For the metric \eqref{eq:gmunu} with power-law coefficients \eqref{eq:power-law}, the Ricci scalar reads
\begin{equation}
    R=-\frac{4 - 4 b + \alpha(2 + \alpha)}{2 b r^2}\,.
\end{equation}
This vanishes only if
\begin{equation}
    \alpha = -1 - \sqrtb{-3 + 4b}\,, \qquad \alpha = -1 + \sqrtb{-3 + 4b}\,,
\end{equation}
Equation \eqref{eq:EOMs} cannot be solved in the former case, but in the latter case one precisely recovers \eqref{eq:solutions}.

\section{Powerballs as cores of quantum black holes}\label{sec:core}

We propose complex powerballs, i.e., the complex metric found in the previous section, as a potential model of the quantum `interior of black holes', which may offer a different understanding of the usual central singularity.
For this to be the case, a complex powerball should resemble a `standard black hole' far away from the central singularity, where in particular GR should hold and where the spacetime is expected to be real and classical far from the black hole's would-be horizon.
By our symmetry assumption on the metric, we will check whether it is possible to `match' a powerball with a Schwarzschild spacetime just outside the Schwarzschild horizon.

Let us start by recalling that the Schwarzschild metric is of the form \eqref{eq:gmunu} with $A(r)=1/B(r)=1-2\GN M/r$. The Schwarzschild horizon would be at $\RS=2\GN M$, so let us consider a radius that is some small $\delta>0$ outside this ($\delta$ has units of length), which we call $r_\star\equiv\RS+\delta$.
For instance, the proper length scale from $\RS$ to $r_\star$ is
\begin{equation}
    \ell_\star\equiv\int\limits_{\RS}^{\RS+\delta}\frac{\dd r}{\sqrtb{1-\RS/r}}=\sqrtb{(\RS+\delta)\delta}+\RS\mathrm{artanh}\left(\sqrtb{\frac{\delta}{\RS+\delta}}\right)\stackrel{\delta\ll\RS}{\simeq}2\sqrtb{\RS\delta}\,,\label{eq:ellstar}
\end{equation}
so for instance this would be of the order of the Planck length (recall $\lp\equiv\sqrtb{\GN}$) if $\delta\sim 1/M$.

We ask for the metric to be described by our complex powerball metric for $|r|\leq r_\star$ and by the Schwarzschild metric for $r\geq r_\star$.
This means gravity is described by the pure quadratic-gravity action \eqref{eq:Spurequad} on scales $|r|\leq r_\star$, while we have Einstein gravity for $r\geq r_\star$.\footnote{This is a similar setup to the original 2-2 hole model \cite{Holdom:2016nek}, although in contrast our power-laws are exact solutions of pure quadratic gravity.} Naturally, a more realistic model would have a proper interpolation between the two regimes, i.e., it is not expected to be a sharp transition at $r_\star$.
For the purpose of this work, we will simply ask for continuity of $g_{tt}$ at $r_\star$ (which ensures the continuity of the induced metric on hypersurfaces of constant radius) and model our ignorance about the nature of gravity in the vicinity of $r_\star$ (more about this in the next section). Continuity of $g_{tt}$ implies
\begin{equation}
    ar_\star^{\alpha_{\mp}}=1-\frac{\RS}{r_\star}\qquad\implies\qquad a=a_\mp\equiv\frac{\delta}{(\RS+\delta)^{1+\alpha_\mp}}\,,\label{eq:match}
\end{equation}
hence this fixes the constant $a$, which we note is complex.

Note that $r^{\alpha_\mp}$ as a complex power is generally multivalued; indeed, we can write $r^\alpha=\exp[\alpha\ln|r|+i\alpha\mathrm{Arg}(r)+2\pi ni\alpha]$ for any $n\in\mathbb{Z}$, where $\mathrm{Arg}(r)$ denotes the principal value of the complex argument of $r$, defined on the interval $(-\pi,\pi]$. When evaluated at the real value $r_\star$, this still means
\begin{equation}
    r_\star^{\alpha_\mp}=e^{2\pi ni\alpha_\mp}\mathrm{PV}[r_\star^{\alpha_\mp}]=e^{n\pi(i\pm\sqrtb{15})}\mathrm{PV}[r_\star^{\alpha_\mp}]=(-1)^ne^{\pm n\pi\sqrtb{15}}\mathrm{PV}[r_\star^{\alpha_\mp}]\,,\label{eq:rstarPV}
\end{equation}
where the principal value (PV) is taken as the $n=0$ case.
Correspondingly, this means that if one goes around the $r=0$ singularity of the powerball in the complex plane (so, e.g., going from $n=0$ to $n=1$), $g_{tt}\propto r_\star^{\alpha_\mp}$ changes sign and acquires an amplification (suppression) factor for the choice of $-$ ($+$) sign. Therefore, if we imagined entering a complex powerball from a Lorentzian spacetime, we would get out on the other side in a Euclidean spacetime at $r=r_\star$. If this journey is repeated, going through a powerball with opposite sign, then one would get back to the same original Lorenztian geometry. We shall elaborate on this journey in the next section.

\section{Complex spacetime geometry representation}\label{sec:geo}

\begin{figure}[t]
    \centering
    \includegraphics[width=0.4\linewidth]{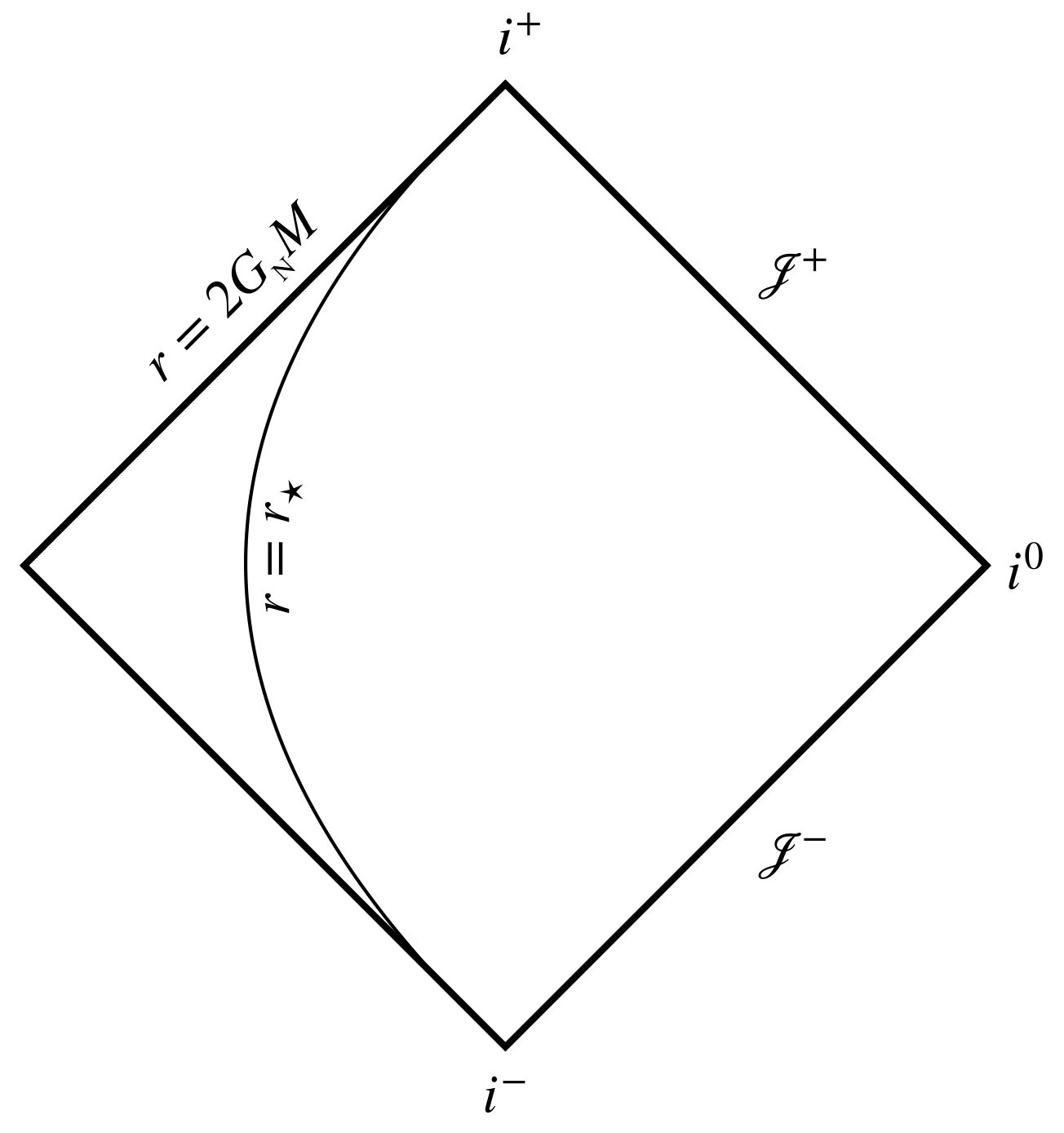}
    \caption{Sketch of the Penrose diagram of an eternal Schwarzschild black hole's exterior geometry. The timelike hypersurface of constant $r=r_\star=2\GN M+\delta$ is shown (not to scale, since it is expected to be about a Planck length above the Schwarzschild radius). Every point in the diagram is a 2-sphere, so the $r=r_\star$ hypersurface is $2+1$ dimensional.}
    \label{fig:SchwPenrose}
\end{figure}

In this section, let us depict the complex space-time geometry of powerballs a little more precisely. The idea is that far away from the powerball interior, the spacetime is the same as that on the exterior of a Schwarzschild black hole. In particular, it is asymptotically flat as represented by the Penrose diagram in Fig.~\ref{fig:SchwPenrose}.\footnote{Here, we are imagining an eternal black hole, not an astrophysical one.}

\begin{figure}[t]
    \centering
    \includegraphics[width=0.6\linewidth]{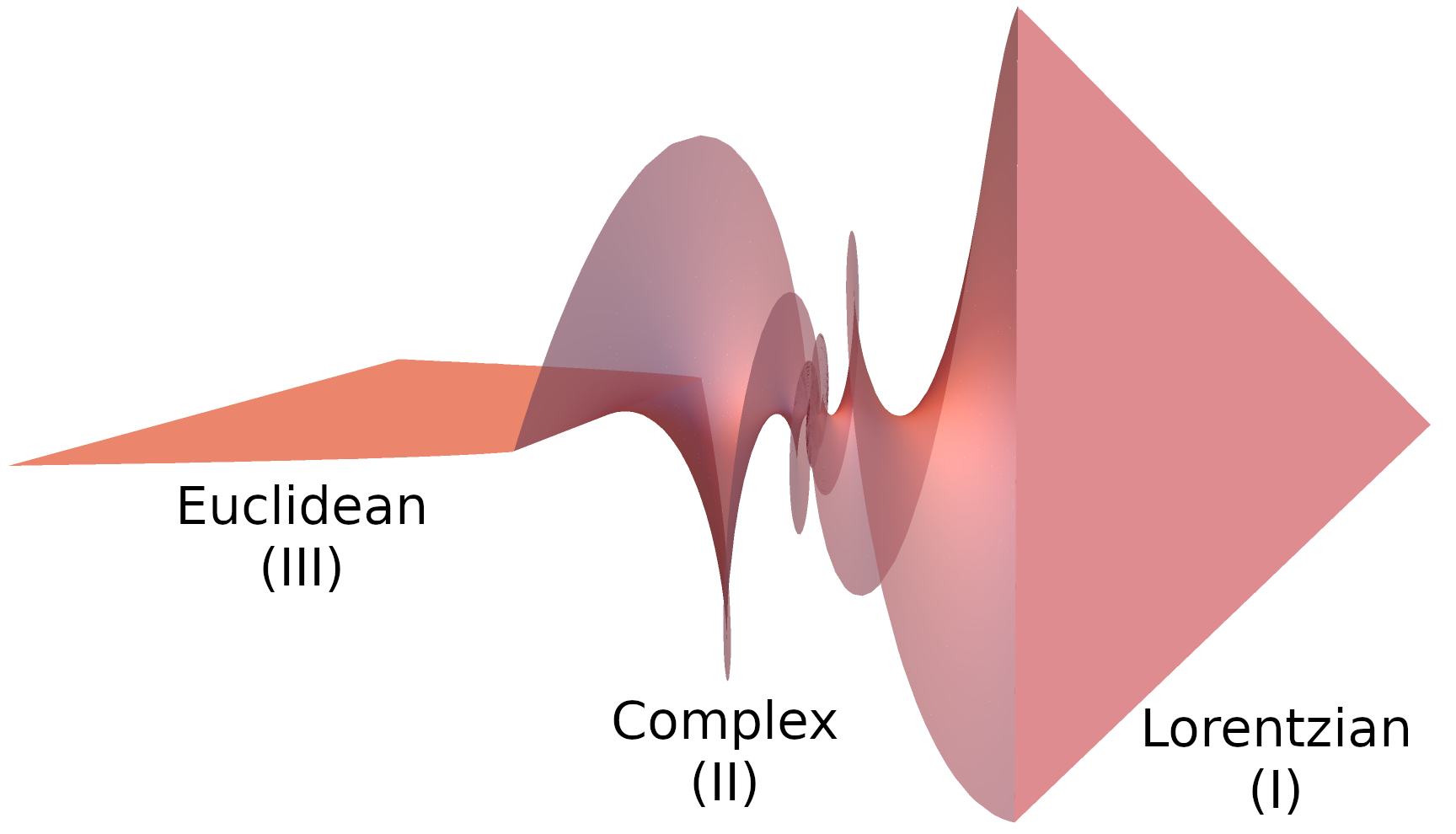}
    \caption{Visual illustration of the complex powerball conformal geometry in the $(r,t)$ plane. The triangle on the right amounts to the Lorentzian Schwarzschild exterior geometry, and what looks like a folded trapezoid on the left constitutes a Euclidean Schwarzschild geometry. The spiraling band in the middle represents the complex powerball geometry. From right to left, the respective regions are labeled as I, II, and III.}
    \label{fig:fullSketch}
\end{figure}

Since we are continuously matching the above Schwarzschild geometry with the complex powerball at $r_\star$, the geometry is no longer real for $|r|\leq r_\star$. Without loss of generality, let us represent the outside geometry simply as a triangle, as if we moved the $r=r_\star$ surface to the vertical axis. Then, we add a dimension to the picture to account for imaginary components. The result is Fig.~\ref{fig:fullSketch}. The illustration highlights the complex nature of $g_{tt}$. When the geometry is Lorentzian, $g_{tt}$ is real and negative [$\mathrm{Arg}(g_{tt})=\pi$], and the corresponding image is in the plane of the page.\footnote{The graph of Fig.~\ref{fig:fullSketch} is slightly rotated and tilted for better visualization.} When the geometry is Euclidean (Riemannian), $g_{tt}$ is real and positive [$\mathrm{Arg}(g_{tt})=0$], and the corresponding image is perpendicular to the plane of the page. Lastly, when $g_{tt}$ has an imaginary part [$\mathrm{Arg}(g_{tt})\in(-\pi,\pi)\setminus\{0\}$], the image has components both in and out of the plane of the page.

Recall that $g_{tt}\propto r^{\alpha_\mp}$ for the powerball metric, where
\begin{equation}
    \Im(\mathrm{PV}[r^{\alpha_\mp}])=\sqrtb{|r|}\exp[\pm\frac{\sqrtb{15}}{2}\mathrm{Arg}(r)]\sin\left[\frac{1}{2}\left(\mathrm{Arg}(r)\mp\sqrtb{15}\ln|r|\right)\right]\,;
\end{equation}
the expression for $\Re(\mathrm{PV}[r^{\alpha_\mp}])$ is the same as the above, except with the sine replaced by a cosine. This is becoming increasingly smaller as $|r|\searrow 0$. At face value, it thus appears as if a singularity is reached at $r=0$; this is the pinching of the geometry in the middle of the complex region in Fig.~\ref{fig:fullSketch}. The spiraling between this pinch and the exterior regions comes from the oscillating functionality, schematically of the form $\sin(\ln|r|)$.

\begin{figure}[t]
    \centering
    \includegraphics[width=0.6\linewidth]{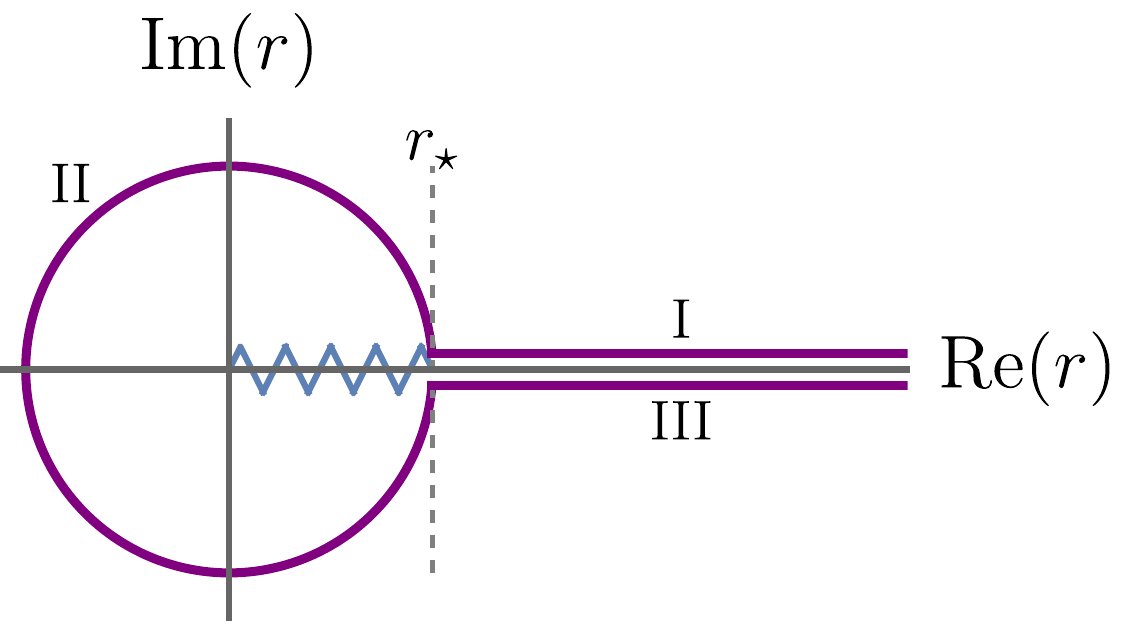}
    \caption{Sketch of the integration contour (purple) in the complex $r$ plane for regions I, II, and III. There is a branch cut (blue) for $0\leq\Re(r)\leq r_\star$ where $\Im(r)=0$.}
    \label{fig:sketchComplexPlane}
\end{figure}

As we argue in the next section, the singularity at $r=0$ is avoidable by going around it in the complex plane from the point of view of an integration contour that avoids the branch cut along the positive real-$r$ axis inside the powerball; see Fig.~\ref{fig:sketchComplexPlane} for a depiction. As we argued below Eq.~\eqref{eq:rstarPV}, this means $g_{tt}(r_\star)$ goes from $a_\mp r_\star^{\alpha_\mp}=\delta/r_\star$ [recall Eq.~\eqref{eq:match}; assuming the principal value of the complex power] to
\begin{equation}
    a_\mp r_\star^{\alpha_\mp}e^{2\pi i\alpha_\mp}=-e^{\pm \pi\sqrtb{15}}\frac{\delta}{r_\star}\,,\label{eq:1rot}
\end{equation}
i.e., it acquires a factor of $e^{2\pi i\alpha_\mp}$ due to the $2\pi$ rotation around the branch cut and singularity. As such, the geometry `rotates' in the complex plane from a Lorentzian signature to a Euclidean one. This is what Fig.~\ref{fig:fullSketch} is meant to convey: the geometry goes from Lorentzian to Euclidean through a spiraling in and out of the powerball.

Interestingly, we note that performing a rotation in the complex plane around the powerball singularity in one direction or another is equivalent to choosing between either $\alpha_+$ or $\alpha_-$. Indeed, since $\alpha_+=\alpha_-^*$ and $\mathrm{Re}(\alpha_\mp)=1/2$, we note that $\alpha_++\alpha_-=1$; hence performing a rotation once with $\alpha_\mp$ and a second one with $\alpha_\pm$ in the same direction takes us back to the starting point:
\begin{equation}
    \frac{\delta}{r_\star}\qquad\stackrel{\circlearrowleft}{\longrightarrow}\qquad\frac{\delta}{r_\star}e^{2\pi i\alpha_\mp}\qquad\stackrel{\circlearrowleft}{\longrightarrow}\qquad\frac{\delta}{r_\star}e^{2\pi i\alpha_\mp}e^{2\pi i\alpha_\pm}=\frac{\delta}{r_\star}\,.
\end{equation}
This is thus equivalent to choosing either $\alpha_+$ or $\alpha_-$ and performing two rotations in opposite directions:
\begin{equation}
    \frac{\delta}{r_\star}\qquad\stackrel{\circlearrowleft}{\longrightarrow}\qquad\frac{\delta}{r_\star}e^{2\pi i\alpha_\mp}\qquad\stackrel{\circlearrowright}{\longrightarrow}\qquad\frac{\delta}{r_\star}e^{2\pi i\alpha_\mp}e^{-2\pi i\alpha_\mp}=\frac{\delta}{r_\star}\,.
\end{equation}
We will see the implications of this from a path integral point of view in the next sections.

\section{Action integral}\label{sec:integral}

\begin{figure}[ht]
  \centering
  \includegraphics[width=0.8\textwidth]{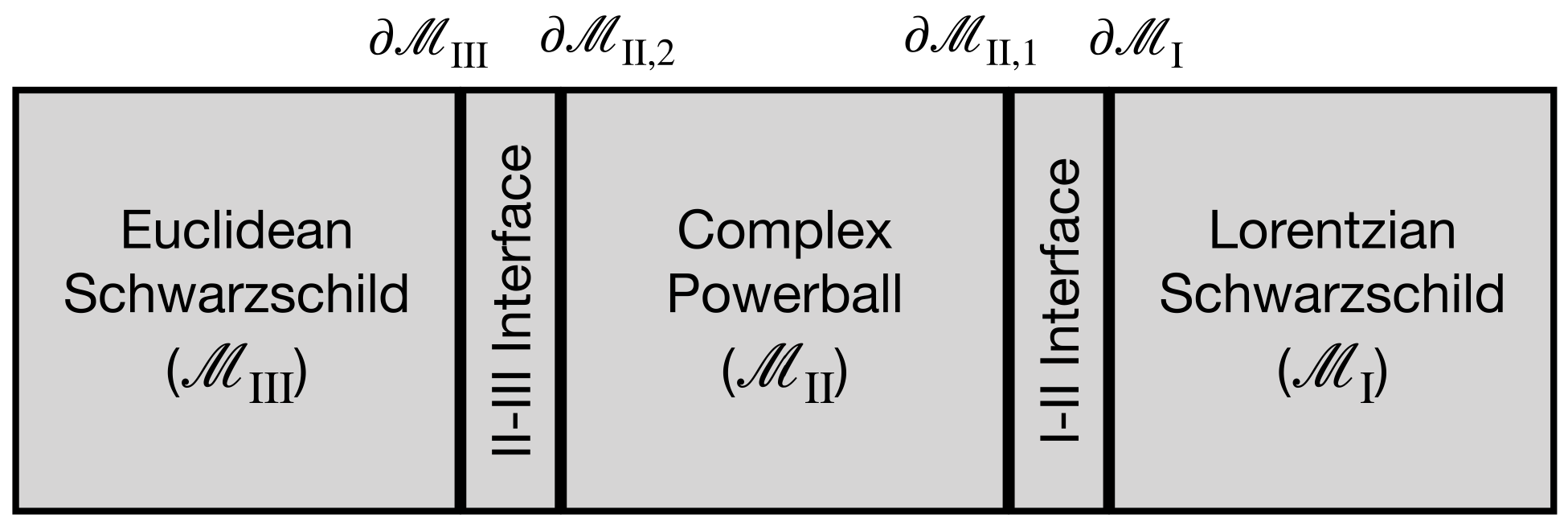}
  \caption{A qualitative cartoon of different regions that contribute to the gravitational action. We depict the I-II and II-III interfaces with some thickness, but in the approximation scheme below we make the transitions between GR and pure quadratic gravity infinitesimally small, such that $\partial\mathcal{M}_\mathrm{I}$ coincides with $\partial\mathcal{M}_\mathrm{II,1}$ and such that $\partial\mathcal{M}_\mathrm{II,2}$ coincides with $\partial\mathcal{M}_\mathrm{III}$.}
  \label{fig:powerball_slide}
\end{figure}

In the spirit of computing a gravitational path integral, let us evaluate the action integral of the whole spacetime under study. Figure \ref{fig:powerball_slide} shows the different regions that contribute to this action. 

\subsection{Lorentzian Schwarzschild}\label{sec:actionLSchw}

To start, recall that for $r\geq r_\star$ we assume to have GR. Let us call this part of the spacetime manifold $\mathcal{M}_\mathrm{I}=\mathbb{R}\times[r_\star,\infty)\times\mathbb{S}^2$, so the total action for this part of spacetime is
\begin{equation}
    S_\textrm{I}=\frac{\mpl^2}{16\pi}\int\limits_{\mathcal{M}_\mathrm{I}}\dd^4x\,\sqrtb{-g}\,R-\frac{\mpl^2}{8\pi}\int\limits_{\partial\mathcal{M}_\mathrm{I}}\dd^3y\,\sqrtb{-h}\,K\,.\label{eq:SI}
\end{equation}
In addition to the Einstein-Hilbert action, note the presence of the Gibbons-Hawking-York (GHY) boundary term due to the boundary of this part of spacetime at $r=r_\star$, denoted as $\mathcal{M}_\mathrm{I}|_{r=r_
\star}=\partial\mathcal{M}_\mathrm{I}=\mathbb{R}\times\mathbb{S}^2$; this is a timelike, $(2+1)$-dimensional Lorentzian hypersurface.
The hypersurface's induced metric is given by
\begin{equation}
    h^{ab}E^\mu_aE^\nu_b=g^{\mu\nu}-n^\mu n^\nu\,,
\end{equation}
where $E^\mu_a=\partial x^\mu/\partial y^a=\delta^\mu_a$ relates the coordinates on $\mathcal{M}_\mathrm{I}$, $x^\mu=\{t,r,\theta,\phi\}$, to those on $\partial\mathcal{M}_\mathrm{I}$, $y^a=\{t,\theta,\phi\}$. Letting $f(r)\equiv\sqrtb{1-\RS/r}$, the spacelike vector $n^\mu=\delta^\mu_rf(r)$ is the unit normal vector to the hypersurface, so $n^\mu n_\mu=1$. Note that $n^\mu$ is pointing in the direction of increasing $r$, which is inwards of $\mathcal{M}_\mathrm{I}$; this is the reason for the minus sign in front of the GHY integral in Eq.~\eqref{eq:SI} --- see, e.g., \cite{Poisson:2009pwt}. Furthermore, $K_{ab}=E^\mu_aE^\nu_b\nabla_\mu n_\nu$ is the extrinsic curvature, and its trace,
\begin{equation}
    K=h^{ab}K_{ab}=\frac{2}{rf(r)}\left(1-\frac{3\RS}{4r}\right)\,,\label{eq:meanCurvSchw}
\end{equation}
is the mean curvature.
The determinant of the induced metric is $h=-f(r)^2r^4\sin^2\theta$, so recalling $R=0$, the action integral reduces to the GHY boundary only at $r_\star$:
\begin{equation}
    S_\textrm{I,on-shell}=-\mpl^2r_\star\left(1-\frac{3\RS}{4r_\star}\right)\Delta t=-\mpl^2\left(\frac{\RS}{4}+\delta\right)\Delta t=-\left(\frac{M}{2}+\mpl^2\delta\right)\Delta t\,;
\end{equation}
for the time being, $\Delta t$ simply denotes the Lorentzian time integration interval.

\subsection{Pure quadratic gravity powerball}

Let us next consider the action integral of the powerball, which consists of pure quadratic gravity for $\mathcal{M}_\mathrm{II}=\mathbb{R}\times\{r\in\mathbb{C}\,\big|\,|r|\leq r_\star\}\times\mathbb{S}^2$ and the corresponding boundary term at $\partial\mathcal{M}_\mathrm{II}$,
\begin{equation}
    S_\textrm{II}=\int\limits_{\mathcal{M}_\mathrm{II}}\dd^4x\,\sqrtb{-g}\left(\frac{\omega}{3\sigma}R^2-\frac{1}{2\sigma}C_{\mu\nu\rho\sigma}^2\right)
    +4\int\limits_{\partial\mathcal{M}_\mathrm{II}}\dd^3y\,\sqrtb{-h}\left(\frac{\omega}{3\sigma}RK+\frac{1}{\sigma}n_\mu n_\nu C^{\mu a\nu b}K_{ab}\right)\,;
\end{equation}
see \cite{Hawking:1984ph,Dyer:2008hb,Deruelle:2009zk,Hohm:2010jc,Jonas:2021xkx} for the boundary term.
Recalling $R=0$ on shell, the action reduces to
\begin{equation}
    S_\textrm{II,on-shell}=-\frac{1}{2\sigma}\int\limits_{\mathcal{M}_\mathrm{II}}\dd^4x\,\sqrtb{-g}\,C_{\mu\nu\rho\sigma}^2
    +\frac{4}{\sigma}\int\limits_{\partial\mathcal{M}_\mathrm{II}}\dd^3y\,\sqrtb{-h}\,n_\mu n_\nu C^{\mu a\nu b}K_{ab}\,.
\end{equation}
Let us call the bulk term $S_\mathrm{powerball}$ and the boundary term $S_{\partial\textrm{powerball}}$, so $S_\textrm{II,on-shell}\equiv S_\mathrm{powerball}+S_{\partial\textrm{powerball}}$.

Let us start by evaluating the powerball bulk term on shell, which recalling Eq.~\eqref{eq:C2s} [and Eqs.~\eqref{eq:gmunu}, \eqref{eq:power-law}, and \eqref{eq:solutions}] amounts to
\begin{equation}
    S_\mathrm{powerball}=-\frac{8}{3\sigma}\int\dd^4x\,\frac{\sqrtb{-g}}{r^4}=-\frac{32\pi\Delta t\sqrtb{a_\mp b_\mp}}{3\sigma}\,\mathcal{I}_\textrm{radial}=-\frac{8\pi\Delta t}{\sigma}\sqrtb{\frac{2a_\mp}{3}\left(1\mp i\sqrtb{15}\right)}\,\mathcal{I}_\textrm{radial}\,,\label{eq:SC2}
\end{equation}
where the radial integral is
\begin{equation}
    \mathcal{I}_\textrm{radial}=\int\limits_{\mathcal{C}}\dd r\frac{\sqrtb{r^{\alpha_\mp}}}{r^2}=\int\limits_{\mathcal{C}}\dd r\,r^{-(7\pm i\sqrtb{15})/4}\,.
\end{equation}
We shall take the contour $\mathcal{C}$ as in Fig.~\ref{fig:sketchComplexPlane} as the path that takes us from $r=r_\star e^{i\epsilon}$, $0<\epsilon\ll 1$, to $r=r_\star e^{i(2\pi-\epsilon)}$, where we eventually want to take the limit $\epsilon\searrow 0$. (Physically, this means the singularity at the center is somehow resolved, and the black hole `bounces', so the path integral takes us from $r_\star$ back to $r_\star$.) To do the integral explicitly, let us parametrize the contour as $r(\vartheta)=r_\star\exp(i\vartheta)$, with $\vartheta\in\mathbb{R}$ ranging from $\epsilon$ to $2\pi-\epsilon$. Then,
\begin{align}
    \mathcal{I}_\textrm{radial}&=\lim_{\epsilon\searrow 0}\int_\epsilon^{2\pi-\epsilon}\dd\vartheta\,i(r_\star e^{i\vartheta})^{1-(7\pm i\sqrtb{15})/4}=\frac{4ir_\star ^{-(3\pm i\sqrtb{15})/4}}{-3i\pm\sqrtb{15}}\lim_{\epsilon\searrow 0}\exp\left[\frac{1}{4}\left(-3i\pm\sqrtb{15}\right)\vartheta\right]_\epsilon^{2\pi-\epsilon}\nonumber\\
    &=~\frac{4r_\star ^{-(3\pm i\sqrtb{15})/4}}{3\pm i\sqrtb{15}}\left(1-ie^{\pm\pi\sqrtb{15}/2}\right)\,.
\end{align}
Substituting the result of this integral into Eq.~\eqref{eq:SC2} and using Eq.~\eqref{eq:match} yields the on-shell powerball bulk action
\begin{equation}
    S_\mathrm{powerball}=\pm\frac{32\pi i\sqrtb{\delta}\Delta t}{3\sigma(\RS+\delta)^{3/2}}\left(1-ie^{\pm\pi\sqrtb{15}/2}\right)\,,\label{qgo}
\end{equation}
which is finite.\footnote{Having a finite on-shell action is a desirable feature from a path integral point of view; see, e.g., \cite{Lehners:2019ibe,Barrow:2019gzc,Jonas:2021xkx,Lehners:2023fud}.} Note that this is the result of a counterclockwise rotation as depicted in Fig.~\ref{fig:sketchComplexPlane}. This is a choice; one more generally finds
\begin{equation}
    S_\mathrm{powerball}=\pm\frac{32\pi i\sqrtb{\delta}\Delta t}{3\sigma(\RS+\delta)^{3/2}}\left(1-i\eta e^{\pm\eta\pi\sqrtb{15}/2}\right)\,,\label{qgo2}
\end{equation}
with $\eta=+1$ for a counterclockwise rotation and $\eta=-1$ for a clockwise rotation.

For the powerball boundary, recalling the powerball metric of Sec.~\ref{sec:powerSol} the timelike hypersurface of constant $r$ corresponding to $\partial\mathcal{M}_\mathrm{II}$ has unit outward-pointing normal vector $n_\mu=\delta_\mu^r\sqrtb{b}$, and the corresponding extrinsic curvature is given by
\begin{equation}
    K_{cd}=\mathrm{diag}\left(-\frac{a\alpha}{2\sqrtb{b}}r^{\alpha-1},\frac{r}{\sqrtb{b}},\frac{r\sin^2\theta}{\sqrtb{b}}\right)\,.
\end{equation}
(The constants $a$, $\alpha$, and $b$ all have $\mp$ values, but we omit the $\mp$ subscript here to lighten the notation.)
Furthermore, one computes
\begin{equation}
    n_\mu n_\nu C^{\mu c\nu d}K_{cd}=-\frac{(\alpha-2)\left((\alpha-2)^2-4b\right)}{24b^{3/2}r^3}=\mp\frac{2i}{3r^3}\,.
\end{equation}
Correspondingly, the powerball boundary term at $r=r_\star$ reduces to
\begin{equation}
    S_{\partial\textrm{powerball}}=-\frac{16\pi}{\sigma}\Delta t\sqrtb{ar_\star^{4+\alpha}}\,\frac{(\alpha-2)\left((\alpha-2)^2-4b\right)}{24b^{3/2}r_\star^3}=\mp\frac{32\pi i\sqrtb{\delta}\Delta t}{3\sigma(\RS+\delta)^{3/2}}\,.\label{eq:dpowerball}
\end{equation}
This corresponds to the boundary term $\partial\mathcal{M}_{\mathrm{II,1}}$, i.e., the part of $\partial\mathcal{M}_\mathrm{II}$ adjacent to $\partial\mathcal{M}_\mathrm{I}$ --- there will be another contribution below in Sec.~\ref{sec:eucSchw} from the part of $\partial\mathcal{M}_\mathrm{II}$ that is adjacent to the Euclidean boundary $\partial\mathcal{M}_\mathrm{III}$. Still, we can already observe that the boundary contribution \eqref{eq:dpowerball} exactly cancels the imaginary part of the bulk action \eqref{qgo2}, and one is only left with a purely real on-shell action for the powerball.

\subsection{Effective interface}

While the induced metric is continuous across the constant-$r_\star$ surface, the extrinsic curvature is not. A proper matching in GR would demand a continuous extrinsic curvature (the Israel junction conditions \cite{Israel:1966rt}), while the condition in quadratic gravity and other modified gravity theories is more complicated (see, e.g., \cite{Reina:2015gxa,Chu:2021uec}). In fact, the transition from GR to pure quadratic gravity should have some thickness, but we assume that it is microscopic in comparison to the macroscopic scales of the system under study. Therefore, we imagine a generic EFT for the transition and collapse its $(3+1)$-dimensional action into an interface action through a delta function that localizes the interface at $\partial\mathcal{M}_\mathrm{I}$. This is akin to a distributional energy-momentum tensor. The resulting (generalized) Israel junction condition, then, would require that this energy-momentum tensor cancels the variation of the boundary terms, inherited from the two sides of the interface, i.e., $S_\textrm{I,on-shell}$ and $S_{\rm \partial powerball}$.

Without proper knowledge of the ``interpolating theory'' from Einstein gravity for $r\geq r_\star$ ($\mathcal{M}_\textrm{I}$) to pure quadratic gravity for $|r|\leq r_\star$ ($\mathcal{M}_\textrm{II}$), we model our ignorance by an effective boundary action for the interface at $r_\star$, i.e., the $(2+1)$ surface of constant $r=r_\star$.\footnote{The same will be true for the transition from regions II to III, but let us focus on the intersection with $\mathcal{M}_\textrm{I}$ for now.}
We take its effective action to be
\begin{equation}
    S_\textrm{interface}^\textrm{I-II}=\frac{\mpl}{8\pi}\int\limits_{\partial\mathcal{M}_\textrm{I}}\dd^3y\,\sqrtb{-h}\left(\zeta\,{}^{(3)}\!R+c_1K^2-c_2K_{ab}^2+\mathcal{O}(\mpl^{-1})\right)\,,\label{eq:interfaceS}
\end{equation}
where ${}^{(3)}\!R=2/r^2$ is the 3-curvature of the constant-$r$ hypersurface, and $\zeta$, $c_1$, and $c_2$ are dimensionless constants. Taking the extrinsic curvature of the boundary hypersurface from the $\mathcal{M}_\textrm{I}$ side as in Sec.~\ref{sec:actionLSchw}, we note that ${}^{(3)}\!R-K^2+K_{ab}^2=0$, in agreement with the scalar Gauss relation (a contracted version of the Gauss-Codazzi relations) for timelike hypersurfaces when the $(3+1)$-dimensional Einstein tensor vanishes (see, e.g., \cite{Poisson:2009pwt}).
Note that a divergence arises if one takes the limit $\delta\searrow 0$ whenever $c_1\neq c_2$ since $r_\star\searrow\RS$ implies, in particular, $f(r)\searrow 0$ and thus $K\to\infty$ [recall Eq.~\eqref{eq:meanCurvSchw}]. If $c_1=c_2\neq 0$, though, the coefficient $\zeta$ in Eq.~\eqref{eq:interfaceS} is simply rescaled according to $\zeta\mapsto\zeta+c_1$ given the scalar Gauss relation ${}^{(3)}\!R=K^2-K_{ab}^2$.\footnote{While we cannot prove that $c_1=c_2$ is the only viable option, it is a necessary condition for the validity of the EFT in the limit $\delta\searrow 0$ to keep the corrections under control. A similar cancellation argument can be made for higher powers of extrinsic curvature, $K^3$, $KK_{ab}^2$, etc.} Therefore, we shall set $c_1=c_2=0$ without loss of generality.

At this point, the action describing the $r=r_\star$ surface between regions I and II has three contributions: the standard GHY action of GR from the Lorentzian Schwarzschild side, the pure quadratic gravity powerball boundary term, and the new interface term we added above. In sum, the total boundary action amounts to
\begin{subequations}
\begin{align}
    S_\mathrm{boundary}^\textrm{I-II}=&\int\limits_{\partial\mathcal{M}_\mathrm{I}}\dd^3y\,\sqrtb{-h}\left(-\frac{\mpl^2}{8\pi}K+\frac{\mpl\zeta}{8\pi}{}^{(3)}\!R\right)+\int\limits_{\partial\mathcal{M}_\mathrm{II,1}}\dd^3y\,\sqrtb{-h}\,\frac{4}{\sigma}n_\mu n_\nu C^{\mu a\nu b}K_{ab}\\
    =&\int\dd^3y\,r^2\sin(\theta)\sqrtb{1-\frac{\RS}{r}}\left(-\frac{\mpl^2}{4\pi r}\frac{1-3\RS/(4r)}{\sqrtb{1-\RS/r}}+\frac{\mpl\zeta}{4\pi r^2}\mp\frac{8i}{3\sigma r^3}\right)\,.
\end{align}
\end{subequations}
Note that intrinsic geometrical quantities such as $h_{ab}$ and ${}^{(3)}\!R$ are continuous across hypersurfaces of constant $r$, which is what allows us to write the above action as a single integral with common measure. By the variational principle, we demand that
\begin{equation}
    \left.\frac{\delta S_\mathrm{boundary}^\textrm{I-II}}{\delta r}\right|_{r=r_\star}=0\,.
\end{equation}
The extremization of the boundary action at $r_\star=\RS+\delta$ thus sets the value $r_\star$ (and correspondingly of $\delta$) in terms of the action's (fundamental) coupling constants. This extremization also essentially amounts to ensuring that we have the appropriate (generalized) Israel junction conditions. In the limit $\delta\ll\RS$, the extremization yields
\begin{equation}
    \delta\simeq\frac{\zeta^2}{4\mpl^2\RS}\mp\frac{16\pi i\zeta}{3\sigma\mpl^3\RS^2}\,.\label{eq:deltazetarel0}
\end{equation}
Assuming $\zeta \sim \sigma \sim {\cal O}(1)$ and a macroscopic black hole $\mpl \RS \gg 1$, the second term is much smaller than the first one, and its imaginary nature is simply representative of the interface as a form of transition from the real Lorentzian Schwarzschild region I to the complex powerball region II. Ignoring this second term, we can write
\begin{equation}
    \delta\approx\frac{\zeta^2}{8M}\,.\label{eq:deltazetarel}
\end{equation}
Therefore, it follows that the interface is a proper distance $\delta$ of the order of the Planck length above the would-be Schwarzschild horizon at $\RS$. Indeed, recalling Eq.~\eqref{eq:ellstar} we can write $\ell_\star\approx\zeta\lp$.

\subsection{Euclidean Schwarzschild}\label{sec:eucSchw}

Recall from the discussion around Eq.~\eqref{eq:1rot} that the complex power-law nature of the powerball metric implies that one goes from a Lorentzian to a Euclidean Schwarzschild geometry in the process of going around the branch cut, from $r_\star$ to $r_\star$. Specifically, as $r$ becomes real again and $r\geq r_\star$, one has the Riemannian manifold $\mathcal{M}_\mathrm{III}=\mathbb{R}\times[r_\star,\infty)\times\mathbb{S}^2$ and metric\footnote{Recall that the choice of $\eta=+1$ or $\eta=-1$ depends on whether the rotation was done in the anticlockwise or clockwise direction, respectively.}
\begin{equation}
    g_{\mu\nu}\dd x^\mu\dd x^\nu=e^{\pm\eta\pi\sqrtb{15}}\left(1-\frac{\RS}{r}\right)\dd t^2+\left(1-\frac{\RS}{r}\right)^{-1}\dd r^2+r^2\dd\Omega^2_{(2)}\,.
\end{equation}
Correspondingly, we have to add yet another boundary action to the whole system given by
\begin{equation}
    S_\mathrm{boundary}^\textrm{II-III}=\int\limits_{\partial\mathcal{M}_\mathrm{II,2}}\dd^3y\,\sqrtb{-h}\,\frac{4}{\sigma}n_\mu n_\nu C^{\mu a\nu b}K_{ab}+\int\limits_{\partial\mathcal{M}_\mathrm{III}}\dd^3y\,\sqrtb{-h}\left(-\frac{\mpl^2}{8\pi}K+\frac{\mpl\zeta}{8\pi}{}^{(3)}\!R\right)\,.
\end{equation}
This accounts for the boundary term of $\mathcal{M}_\mathrm{II}$ at the intersection with $\mathcal{M}_\mathrm{III}$, the transition interface between the two regions (with the same coupling constant $\zeta$ as before), and the GHY term of the Euclidean Schwarzschild region III. Since we are interested in the on-shell action, we omit the vanishing bulk action of region III.
A similar calculation to what was outlined in the previous subsections thus yields the on-shell action
\begin{equation}
    S_\mathrm{boundary}^\textrm{II-III}=i \eta e^{\pm\eta\pi\sqrtb{15}/2}\Delta t\left(-\mpl^2\left(\frac{\RS}{4}+\delta\right)+\mpl\zeta\sqrtb{\frac{\delta}{\RS+\delta}}\mp\frac{32\pi i}{3\sigma}\frac{\sqrtb{\delta}}{(\RS+\delta)^{3/2}}\right)\,.
\end{equation}
Note that $S_\mathrm{boundary}^\textrm{I-II}$ is the same as the above expression for $S_\mathrm{boundary}^\textrm{II-III}$, except for a prefactor of $i \eta e^{\pm\eta\pi\sqrtb{15}/2}$. Further note that the $\eta=+1$ counterclockwise rotation case is akin to an anti-Wick rotation where $t\mapsto e^{\pm\pi\sqrtb{15}/2}it$, while the $\eta=-1$ clockwise rotation case is akin to a Wick rotation where $t\mapsto -e^{\mp\pi\sqrtb{15}/2}it$.

\subsection{Total action}

Gathering all of the action integrals derived above, we find the total on-shell action to be
\begin{align}
    S_\textrm{total}^\textrm{on-shell}=&~S_\mathrm{boundary}^\textrm{I-II}+S_\textrm{powerball}+S_\mathrm{boundary}^\textrm{II-III}\nonumber\\
    \approx &-\left(1+i\eta e^{\pm\eta\pi\sqrtb{15}/2}\right)\left(1-\frac{\zeta^2\mpl^2}{4M^2}\right)\frac{M \Delta t}{2}\pm\eta\frac{8\pi e^{\pm\eta\pi\sqrtb{15}/2}\zeta\mpl^3\Delta t}{3\sigma M^2}\,,
\end{align}
where the approximation uses the relation \eqref{eq:deltazetarel0} between $\delta$ and $\zeta$ and assumes $M\gg\mpl$.
We will shortly be interested in the quantity $\exp(iS_\textrm{total}^\textrm{on-shell})=\exp(\mathcal{W}+i\mathcal{S})$, where
\begin{subequations}
\begin{align}
    \mathcal{W}&\equiv\Re(iS_\textrm{total}^\textrm{on-shell})=\eta e^{\pm\eta\pi\sqrtb{15}/2}\left[1-\frac{\zeta^2\mpl^2}{4M^2}+\mathcal{O}\Big(\frac{\mpl^4}{M^4}\Big)\right]\frac{M\Delta t}{2}\,,\label{eq:Re}\\
    \mathcal{S}&\equiv\Im(iS_\textrm{total}^\textrm{on-shell})=-\left[1-\frac{\zeta^2\mpl^2}{4M^2}\pm\eta\frac{16\pi e^{\pm\eta\pi\sqrtb{15}/2}\zeta\mpl^3}{3\sigma M^3}+\mathcal{O}\Big(\frac{\mpl^4}{M^4}\Big)\right]\frac{M\Delta t}{2}\,.\label{eq:Im}
\end{align}
\end{subequations}
The former is called the weight and the latter the phase.

\section{Physical interpretation of the wave function}\label{sec:path}

\subsection{Lorentzian and Euclidean path integrals}

One can make sense of the complex powerball geometry from a quantum gravity point of view. Indeed, an approach to quantum gravity is to define the wave function of the Universe as the path integral over all possible spacetime geometries (i.e., metrics):
\begin{equation}
    \Psi=\int\mathcal{D}\bm{g}\,e^{iS[\bm{g}]}\,.
\end{equation}
In a semiclassical sense, the path integral will be dominated by a sum over all possible saddle-point `instanton' geometries, i.e., metrics (not necessarily real) that solve the classical equations of motion of the theory.

In Lorentzian quantum cosmology, the wave function can be thought of as a transition amplitude between two spacelike hypersurfaces, call them $\Sigma_{\mathrm{in}}$ and $\Sigma_{\mathrm{out}}$, so
\begin{equation}
    \Psi[\Sigma_{\mathrm{in}}\to\Sigma_{\mathrm{out}}]=\langle\Sigma_{\mathrm{out}}|\Sigma_{\mathrm{in}}\rangle=\int\limits_{\Sigma_{\mathrm{in}}}^{\Sigma_{\mathrm{out}}}\mathcal{D}\bm{g}\,e^{iS[\bm{g}]}\sim\exp(iS_\textrm{on-shell}[\Sigma_{\mathrm{in}}\to\Sigma_{\mathrm{out}}])\,,
\end{equation}
where the last equality indicates the saddle-point approximation: the wave function will be dominated by the on-shell action, i.e., the action evaluated on a solution to the classical equations of motion that satisfy the boundary conditions at $\Sigma_{\mathrm{in}}$ and $\Sigma_{\mathrm{out}}$; we further omit a normalization constant. Such a formulation of the wave function has a probabilistic interpretation for the transition from $\Sigma_{\mathrm{in}}$ to $\Sigma_{\mathrm{out}}$ given by (see, e.g., \cite{Lehners:2023yrj})
\begin{equation}
    \mathrm{Prob}[\Sigma_{\mathrm{in}}\to\Sigma_{\mathrm{out}}]\sim\big|\Psi[\Sigma_{\mathrm{in}}\to\Sigma_{\mathrm{out}}]\big|^2\sim\exp(2\mathcal{W})\,,\qquad\mathcal{W}\equiv\Re(iS_\textrm{on-shell}[\Sigma_{\mathrm{in}}\to\Sigma_{\mathrm{out}}])\,.
\end{equation}
Note that this only yields a relative probability density.

If applied to the results of the previous section, the above tells us that \eqref{eq:Im} only contributes a phase to the wave function, while \eqref{eq:Re} suggests
\begin{equation}
    \mathrm{Prob}\sim\exp(\eta e^{\pm\eta\pi\sqrtb{15}/2}\left(1-\frac{\zeta^2\mpl^2}{4M^2}\right)M\Delta t)\,. \label{eq:prob_Lorentz}
\end{equation}
The interpretation differs drastically depending on the choice of $\eta=\pm 1$, i.e., the direction of the complex contour integral.\footnote{This is similar to the two interpretations of the no-boundary scenario in quantum cosmology, where the Hartle-Hawking \cite{Hartle:1983ai} and the Vilenkin \cite{Vilenkin:1982de,Vilenkin:1983xq,Vilenkin:1984wp} wave functions have weights of opposite sign, which has important implications about their stability and likeliness; see, e.g., \cite{Linde:1983mx,Feldbrugge:2017fcc,Lehners:2023yrj}.} In the case of a `standard' Wick rotation (so with $\eta=-1$), the probability is exponentially suppressed with time (at fixed $\zeta\mpl/M\ll 1$), suggesting that powerballs are virtual objects that are unstable and quickly tend to `decay' over time, in accordance with Heisenberg's uncertainty principle $\Delta E\, \Delta t \gtrsim 1/2$. However, if we treat $\zeta$ as a parameter that can vary across the Hilbert space, its probability is unbounded, with large values of $\zeta$ exponentially favored, going beyond the regime of validity of the approximations made. 

Alternatively, in the case of an anti-Wick rotation (so with $\eta=+1$) and at fixed $\zeta$, the probability is exponentially enhanced with time, suggesting powerballs may be the preferred end point of gravitational collapse. With the same choice of $\eta$, a probabilistic interpretation of $\zeta$ favors a Gaussian centered at $\zeta=0$, with a variance of ${\cal O} (M \Delta t/\mpl)$. 

Another interpretation of the wave function is in terms of a finite-temperature partition function \cite{Gibbons:1976ue}, which is related to the Euclidean path integral as
\begin{equation}
    Z=\Tr e^{-\beta\hat{H}}=\langle\tau+\beta|e^{-\beta\hat H}|\tau\rangle=\int\limits_\tau^{\tau+\beta}\mathcal{D}\bm{g}\,e^{-S_\mathrm{E}[\bm{g}]}\implies\ln|Z|\approx-\Re(S_\mathrm{E}^\textrm{on-shell})\,,
\end{equation}
where $\beta=T^{-1}$ is the inverse temperature, $\hat H$ is the Hamiltonian corresponding to the Euclidean action $S_\mathrm{E}$, and $\tau$ is the Euclidean time.
This is related to the Lorentzian path integral by performing the Wick rotation $t\mapsto -i\tau$ (hence we choose $\eta=-1$ henceforth) and periodically identifying the Euclidean time $\tau\cong\tau+\Delta\tau$ with the inverse temperature $\Delta\tau=\beta$. In our case, Eq.~\eqref{eq:Im} implies
\begin{equation}
    \ln|Z|\approx-\Re(S_\mathrm{E}^\textrm{on-shell})\approx-\frac{M\beta}{2}\left(1-\frac{\zeta^2\mpl^2}{4M^2}\mp\frac{16\pi e^{\mp\pi\sqrtb{15}/2}\zeta\mpl^3}{3\sigma M^3}\right)\,.
\end{equation}
While we cannot identify $\beta$ with $4\pi\RS$ as in the usual Schwarzschild geometry since we are `cutting' it at $r_\star=\RS+\delta$, there nevertheless are arguments pointing to the fact that compact not-quite-black-hole (horizonless) objects should have the same thermodynamic properties as black holes \cite{Mathur:2024mvo}. Therefore, let us take $\beta=1/T_\mathrm{Hawk}$ with $T_\mathrm{Hawk}=\mpl^2/(8\pi M)$, which should hold to leading order.
This implies the entropy is given by
\begin{equation}
    \mathsf{S}=(1-\beta\partial_\beta)\ln|Z|\approx\frac{4\pi M^2}{\mpl^2}\left(1+\frac{\zeta^2\mpl^2}{4M^2}\pm\frac{32\pi e^{\mp\pi\sqrtb{15}/2}\zeta\mpl^3}{3\sigma M^3}\right)\,.
\end{equation}
This is consistent with the area law of a `black hole' of size $r_\star$ to leading order.

\subsection{Allowance of the metric}

Since the powerball metric is complex in nature, one may question its validity and viability, including in the context of a gravitational path integral. 
Motivated in part by a more formal and axiomatic formulation of quantum field theory in curved spacetime, Kontsevich \& Segal \cite[KS]{Kontsevich:2021dmb} showed that demanding arbitrary real $p$-form gauge fields to have convergent path integrals on a fixed metric $g_{\mu\nu}$ leads to a criterion on how complex the metric can be. A simpler criterion, though of the same form (found by demanding convergent scalar field path integrals), had already been found by Louko \& Sorkin \cite{Louko:1995jw}. It was realized by Witten \cite{Witten:2021nzp} that the KS criterion could be used as a mean of determining which complex metrics are to be allowed in a gravitational path integral, as opposed to which complex metrics should be discarded since they would lead to instabilities. Applied in such a way, the KS criterion has led to many interesting results regarding the allowability of complex metrics, especially in the context of quantum cosmology (see, e.g., \cite{Lehners:2021mah,Jonas:2022uqb,Lehners:2022xds,Hertog:2023vot,Lehners:2023pcn,Maldacena:2024uhs,Janssen:2024vjn,Chakravarty:2024bna,Hertog:2024nbh}).

For a general complex metric that may be diagonalizable in a Euclidean basis, so $g_{\mu\nu}=\lambda_{(\mu)}\delta_{\mu\nu}$, the KS criterion can be written as a bound on the amplitude of the sum of the complex argument of each diagonal element (see, e.g., \cite{Kontsevich:2021dmb,Witten:2021nzp,Lehners:2021mah}),
\begin{equation}
    \sum_{\mu=0}^3\Big|\mathrm{Arg}\left[\lambda_{(\mu)}\right]\!\Big|<\pi\,,\label{eq:KS}
\end{equation}
and this inequality should hold at every point in spacetime.\footnote{Saturation of the inequality implies conditional convergence of matter path integrals, and this is what happens for purely real Lorentzian metrics (see, e.g., \cite{Jonas:2022uqb,Feldbrugge:2017kzv}).}
In the on-shell action derived in the previous section (which then enters the path integral as explored above), the powerball metric can be expressed as
\begin{equation}
    g_{\mu\nu}\dd x^\mu\dd x^\nu=-ar(\vartheta)^\alpha\dd t^2+br'(\vartheta)^2\dd\vartheta^2+r(\vartheta)^2\dd\Omega_{(2)}^2\,,
\end{equation}
where a prime denotes a derivative with respect to the argument of the function.
In computing the action integral, we took $r(\vartheta)=r_\star e^{i\eta\vartheta}$ for simplicity, but the contour could generally be deformed (at fixed end points and without crossing the branch cut). One question we could ask is whether there exists a complex integration contour, i.e., a function $r(\vartheta):\mathbb{R}\to\mathbb{C}$ starting and ending at $r_\star$, such that the KS bound \eqref{eq:KS} is respected everywhere along the path, meaning
\begin{equation}
    \Big|\mathrm{Arg}\left[-ar(\vartheta)^\alpha\right]\!\Big|+\Big|\mathrm{Arg}\left[br'(\vartheta)^2\right]\!\Big|
    +2\Big|\mathrm{Arg}\left[r(\vartheta)^2\right]\!\Big|<\pi\label{eq:KSp}
\end{equation}
for all $\vartheta$.
This is clearly not the case for $r(\vartheta)=r_\star e^{i\eta\vartheta}$ since, for instance at $\vartheta=\pi/2$, the left-hand side of \eqref{eq:KSp} is already greater or equal to $2|\mathrm{Arg}[r(\pi/2)^2]|=2\pi$, hence the bound is violated. In fact, it looks like this is inevitable: any path that goes around the branch cut will have a point where $\mathrm{Arg}(r)=\pi/2$, hence where the left-hand side of \eqref{eq:KSp} will be $\geq 2\pi$.

Similar `singularity resolving' complex integration contours in the context of cosmology violate the KS criterion\footnote{This is not always true either; see, e.g., \cite{Chakravarty:2024bna}.} by a similar argument (see, e.g., \cite{Jonas:2022uqb}), but we should mention a few caveats. While the KS criterion is physically reasonable and often leads to sensible conclusions, it can also be too constraining at times (see, e.g., \cite{Chen:2023hra}); in particular, the theory might not admit arbitrary real $p$-form gauge fields to start with (see \cite{Lehners:2022xds,Hertog:2023vot,Lehners:2023pcn,Maldacena:2024uhs,Janssen:2024vjn,Hertog:2024nbh} for works with a complex scalar field). In such a case, the criterion can be weakened.

In fact, the premise of this work has been that, while pure quadratic gravity is generally not quite conformally invariant because of its $R^2$ term, the theory should be conformally invariant near would-be singularities, e.g., be pure Weyl squared gravity when the background metric is Ricci flat. In such a case, coupling arbitrary $p$-form gauge fields to Weyl gravity would break conformal invariance, so it would appear unreasonable to do so and demand stability of such matter fields. Instead, one should ask for the convergence of matter that can be added to the theory in a conformally invariant way, e.g., a real scalar with action
\begin{equation}
    S_\phi=\int\dd^4x\,\sqrtb{-g}\Big((\nabla^\mu\nabla_\mu\phi)^2-2\Big(R_{\mu\nu}-\frac{1}{3}Rg_{\mu\nu}\Big)\nabla^\mu\phi\nabla^\nu\phi
    +F(\phi)C_{\mu\nu\rho\sigma}^2\Big)\,,
\end{equation}
for any real function $F(\phi)$. Indeed, $S_\phi$ is invariant under $g_{\mu\nu}\mapsto\Omega^2g_{\mu\nu}$ and $\phi\mapsto\phi$ for any spacetime-dependent function $\Omega$. The condition on $g_{\mu\nu}$ for the convergence of $\int\mathcal{D}\phi\,\exp(iS_\phi[\phi,g_{\mu\nu}])$ is not known (and most likely highly nontrivial); certainly, there is no reason for it to correspond to the KS criterion. We do not explore this avenue further in this work; it could be an interesting followup question, including, more generally, how matter fields affect powerballs. We end by noting that there is a parallel with proposals that suggest conformal invariance should play a role in the very early Universe when it comes to the cosmological `big bang' singularity (see, e.g., \cite{Bars:2013yba,Bars:2013vba,Boyle:2018tzc,Boyle:2021jej,Boyle:2021jaz,Turok:2022fgq,Boyle:2022lcq,Turok:2023amx}).

\section{Conclusions}\label{sec:concl}

In this work, we explored properties of powerballs, complex power-law solutions to pure quadratic gravity that could represent the interior of quantum black holes since they admit a continuous matching to a Schwarzschild exterior spacetime. We modeled the transition from GR to pure quadratic gravity as an effective interface action, whose extremization set the distance of the interface about a Planck length from the would-be Schwarzschild horizon. We interpreted the complex geometry as a gravitational instanton, which involved computing the powerball on-shell action with a choice of integration contour in the complex plane that avoids the solution's branch cut and singularity. As a result of the rotation in the complex plane, powerballs are Lorentzian on one side and Euclidean on the other. Depending on the direction of the rotation (akin to choosing a Wick rotation), the wave function interpreted as a transition amplitude implies a `tunneling' probability, either corresponding to unstable virtual objects or stable solitons. With the standard Wick rotation, we also estimated the thermal partition function, which reproduces the thermodynamic properties of GR black holes up to small corrections.

The work could be extended in several directions, e.g., including charge (cf.~Reissner-Nordstr\"om; this might allow us to better understand the thermal partition function, the entropy, and the dependence on temperature) or going beyond spherical symmetry to allow for perturbations (e.g., quantum fluctuations). Also, while we treated $\zeta$ as a fixed parameter of the `interface theory', we briefly mentioned a probabilistic interpretation of $\zeta$ (in which case it would not be a fixed parameter of the interface action), which could resonate with ensembles of theories in the context of gravitational path integrals (e.g., \cite{Marolf:2020xie,VanRaamsdonk:2020tlr}).

Other interesting questions include what lies beyond the Euclidean region, for instance whether one could `re-enter' a powerball and `reexit' in a Lorentzian region again. A better understanding of the overall causal spacetime development would certainly be of interest. Since powerballs have two solutions, with constants $\alpha_\mp,b_\mp,a_\mp$, it would also be interesting to explore whether multipowerball solutions are possible and what would be the implications at the level of the wave function. Could this lead to some interference between powerball solutions of the form $\Psi\sim e^{iS_-}+e^{iS_+}$?
Moreover, it would be interesting to drop the starting assumption that running of the coupling constants is negligible; indeed the theory must run into GR in the infrared. How would RG flow affect the powerball solutions and what would be the implications? Perhaps this could also help us improve our understanding and develop a smoother description of the transition from the GR exterior to the conformal core. Indeed, while we found an exact solution to pure classical quadratic gravity and the Schwarzschild metric is an exact solution to pure GR, both are likely only approximate solutions to the complete quantum theory, i.e., full quadratic gravity with running of its couplings under RG flow.

Finally, it is important to explore the potential observable consequences of complex powerballs for astrophysical and gravitational wave observations, as ultimately it is empirical data that will test our theoretical fantasies.

\vskip23pt
\subsection*{Acknowledgments}
We would like thank  John Donoghue, Bob Holdom, Samir Mathur, and Jing Ren for useful discussions.  
This research is supported in part by the Perimeter Institute for Theoretical Physics and the National Sciences and Engineering Research Council of Canada (NSERC).
Research at the Perimeter Institute is supported by the Government of Canada through the Department of Innovation, Science and Economic Development and by the Province of Ontario through the Ministry of Colleges and Universities.
J.Q.~further acknowledges financial support from the University of Waterloo's Faculty of Mathematics William T.~Tutte Postdoctoral Fellowship.
\vskip23pt

\vskip23pt
\subsection*{Data availability}
No data were created or analyzed in this study.
\vskip23pt

%%% APPENDICES %%%

%\appendix

%\section{}

%%% BIBLIOGRAPHY %%%

%\newpage % can be commented out depending on spacing of previous sections and footnotes therein
%\phantomsection
\addcontentsline{toc}{section}{References}

% to remove spacing between items and make the font smaller
\let\oldbibliography\thebibliography
\renewcommand{\thebibliography}[1]{
  \oldbibliography{#1}
  \setlength{\itemsep}{0pt}
  \footnotesize % \small by default
}

\bibliographystyle{JHEP2}
\bibliography{refs}

\providecommand{\url}[1]{#1}\providecommand{\href}[2]{#2}\begingroup\raggedright\begin{thebibliography}{100}

\bibitem{Penrose:1964wq}
R.~Penrose, \emph{{Gravitational collapse and space-time singularities}},
  \href{https://doi.org/10.1103/PhysRevLett.14.57}{\emph{Phys. Rev. Lett.}
  {\bfseries 14} (1965) 57}.

\bibitem{Donoghue:1994dn}
J.F.~Donoghue, \emph{{General relativity as an effective field theory: The
  leading quantum corrections}},
  \href{https://doi.org/10.1103/PhysRevD.50.3874}{\emph{Phys. Rev. D}
  {\bfseries 50} (1994) 3874}
  [\href{https://arxiv.org/abs/gr-qc/9405057}{{\ttfamily gr-qc/9405057}}].

\bibitem{Donoghue:1995cz}
J.F.~Donoghue, \emph{{Introduction to the effective field theory description of
  gravity}},  in \emph{{Advanced School on Effective Theories}}, 1995,
  \href{https://arxiv.org/abs/gr-qc/9512024}{{\ttfamily gr-qc/9512024}}.

\bibitem{Burgess:2003jk}
C.P.~Burgess, \emph{{Quantum gravity in everyday life: General relativity as an
  effective field theory}},
  \href{https://doi.org/10.12942/lrr-2004-5}{\emph{Living Rev. Rel.} {\bfseries
  7} (2004) 5} [\href{https://arxiv.org/abs/gr-qc/0311082}{{\ttfamily
  gr-qc/0311082}}].

\bibitem{Donoghue:2012zc}
J.F.~Donoghue, \emph{{The effective field theory treatment of quantum
  gravity}}, \href{https://doi.org/10.1063/1.4756964}{\emph{AIP Conf. Proc.}
  {\bfseries 1483} (2012) 73}
  [\href{https://arxiv.org/abs/1209.3511}{{\ttfamily arXiv:1209.3511}}].

\bibitem{Donoghue:2017pgk}
J.F.~Donoghue, M.M.~Ivanov and A.~Shkerin, \emph{{EPFL Lectures on General
  Relativity as a Quantum Field Theory}},
  \href{https://arxiv.org/abs/1702.00319}{{\ttfamily arXiv:1702.00319}}.

\bibitem{Ruhdorfer:2019qmk}
M.~Ruhdorfer, J.~Serra and A.~Weiler, \emph{{Effective Field Theory of Gravity
  to All Orders}}, \href{https://doi.org/10.1007/JHEP05(2020)083}{\emph{JHEP}
  {\bfseries 05} (2020) 083}
  [\href{https://arxiv.org/abs/1908.08050}{{\ttfamily arXiv:1908.08050}}].

\bibitem{Donoghue:2022eay}
J.F.~Donoghue, \emph{{Quantum General Relativity and Effective Field Theory}},
  in \emph{{Handbook of Quantum Gravity}} (C.~Bambi, L.~Modesto and
  I.L.~Shapiro, eds.), 2023, \href{https://arxiv.org/abs/2211.09902}{{\ttfamily
  arXiv:2211.09902}},
  \href{https://doi.org/10.1007/978-981-19-3079-9_1-1}{DOI}.

\bibitem{tHooft:1974toh}
G.~'t~Hooft and M.J.G.~Veltman, \emph{{One loop divergencies in the theory of
  gravitation}}, {\emph{Ann. Inst. H. Poincare A Phys. Theor.} {\bfseries 20}
  (1974) 69}, \url{http://www.numdam.org/item/AIHPA_1974__20_1_69_0/}.

\bibitem{Christensen:1979iy}
S.M.~Christensen and M.J.~Duff, \emph{{Quantizing Gravity with a Cosmological
  Constant}}, \href{https://doi.org/10.1016/0550-3213(80)90423-X}{\emph{Nucl.
  Phys. B} {\bfseries 170} (1980) 480}.

\bibitem{Goroff:1985th}
M.H.~Goroff and A.~Sagnotti, \emph{{The Ultraviolet Behavior of Einstein
  Gravity}}, \href{https://doi.org/10.1016/0550-3213(86)90193-8}{\emph{Nucl.
  Phys. B} {\bfseries 266} (1986) 709}.

\bibitem{Schwarz:1982jn}
J.H.~Schwarz, \emph{{Superstring Theory}},
  \href{https://doi.org/10.1016/0370-1573(82)90087-4}{\emph{Phys. Rept.}
  {\bfseries 89} (1982) 223}.

\bibitem{Ashtekar:1986yd}
A.~Ashtekar, \emph{{New Variables for Classical and Quantum Gravity}},
  \href{https://doi.org/10.1103/PhysRevLett.57.2244}{\emph{Phys. Rev. Lett.}
  {\bfseries 57} (1986) 2244}.

\bibitem{Rovelli:1994ge}
C.~Rovelli and L.~Smolin, \emph{{Discreteness of area and volume in quantum
  gravity}}, \href{https://doi.org/10.1016/0550-3213(95)00150-Q}{\emph{Nucl.
  Phys. B} {\bfseries 442} (1995) 593}
  [\href{https://arxiv.org/abs/gr-qc/9411005}{{\ttfamily gr-qc/9411005}}].

\bibitem{Rovelli:1997yv}
C.~Rovelli, \emph{{Loop quantum gravity}},
  \href{https://doi.org/10.12942/lrr-1998-1}{\emph{Living Rev. Rel.} {\bfseries
  1} (1998) 1} [\href{https://arxiv.org/abs/gr-qc/9710008}{{\ttfamily
  gr-qc/9710008}}].

\bibitem{Ashtekar:2004eh}
A.~Ashtekar and J.~Lewandowski, \emph{{Background independent quantum gravity:
  A Status report}},
  \href{https://doi.org/10.1088/0264-9381/21/15/R01}{\emph{Class. Quant. Grav.}
  {\bfseries 21} (2004) R53}
  [\href{https://arxiv.org/abs/gr-qc/0404018}{{\ttfamily gr-qc/0404018}}].

\bibitem{Bombelli:1987aa}
L.~Bombelli, J.~Lee, D.~Meyer and R.~Sorkin, \emph{{Space-Time as a Causal
  Set}}, \href{https://doi.org/10.1103/PhysRevLett.59.521}{\emph{Phys. Rev.
  Lett.} {\bfseries 59} (1987) 521}.

\bibitem{Sorkin:2003bx}
R.D.~Sorkin, \emph{{Causal sets: Discrete gravity}},  in \emph{{School on
  Quantum Gravity}}, 2003,
  \href{https://arxiv.org/abs/gr-qc/0309009}{{\ttfamily gr-qc/0309009}},
  \href{https://doi.org/10.1007/0-387-24992-3_7}{DOI}.

\bibitem{Dowker:2005tz}
F.~Dowker, \emph{{Causal sets and the deep structure of spacetime}},  in
  \emph{{100 Years Of Relativity}: {space-time structure: Einstein and beyond}}
  (A.~Ashtekar, ed.), 2005,
  \href{https://arxiv.org/abs/gr-qc/0508109}{{\ttfamily gr-qc/0508109}}.

\bibitem{Surya:2019ndm}
S.~Surya, \emph{{The causal set approach to quantum gravity}},
  \href{https://doi.org/10.1007/s41114-019-0023-1}{\emph{Living Rev. Rel.}
  {\bfseries 22} (2019) 5} [\href{https://arxiv.org/abs/1903.11544}{{\ttfamily
  arXiv:1903.11544}}].

\bibitem{Ambjorn:2005db}
J.~Ambjorn, J.~Jurkiewicz and R.~Loll, \emph{{Spectral dimension of the
  universe}}, \href{https://doi.org/10.1103/PhysRevLett.95.171301}{\emph{Phys.
  Rev. Lett.} {\bfseries 95} (2005) 171301}
  [\href{https://arxiv.org/abs/hep-th/0505113}{{\ttfamily hep-th/0505113}}].

\bibitem{Ambjorn:1998xu}
J.~Ambjorn and R.~Loll, \emph{{Nonperturbative Lorentzian quantum gravity,
  causality and topology change}},
  \href{https://doi.org/10.1016/S0550-3213(98)00692-0}{\emph{Nucl. Phys. B}
  {\bfseries 536} (1998) 407}
  [\href{https://arxiv.org/abs/hep-th/9805108}{{\ttfamily hep-th/9805108}}].

\bibitem{Ambjorn:2012jv}
J.~Ambjorn, A.~Goerlich, J.~Jurkiewicz and R.~Loll, \emph{{Nonperturbative
  Quantum Gravity}},
  \href{https://doi.org/10.1016/j.physrep.2012.03.007}{\emph{Phys. Rept.}
  {\bfseries 519} (2012) 127}
  [\href{https://arxiv.org/abs/1203.3591}{{\ttfamily arXiv:1203.3591}}].

\bibitem{Loll:2019rdj}
R.~Loll, \emph{{Quantum Gravity from Causal Dynamical Triangulations: A
  Review}}, \href{https://doi.org/10.1088/1361-6382/ab57c7}{\emph{Class. Quant.
  Grav.} {\bfseries 37} (2020) 013002}
  [\href{https://arxiv.org/abs/1905.08669}{{\ttfamily arXiv:1905.08669}}].

\bibitem{Niedermaier:2006wt}
M.~Niedermaier and M.~Reuter, \emph{{The Asymptotic Safety Scenario in Quantum
  Gravity}}, \href{https://doi.org/10.12942/lrr-2006-5}{\emph{Living Rev. Rel.}
  {\bfseries 9} (2006) 5}.

\bibitem{Codello:2008vh}
A.~Codello, R.~Percacci and C.~Rahmede, \emph{{Investigating the Ultraviolet
  Properties of Gravity with a Wilsonian Renormalization Group Equation}},
  \href{https://doi.org/10.1016/j.aop.2008.08.008}{\emph{Annals Phys.}
  {\bfseries 324} (2009) 414}
  [\href{https://arxiv.org/abs/0805.2909}{{\ttfamily arXiv:0805.2909}}].

\bibitem{Benedetti:2009rx}
D.~Benedetti, P.F.~Machado and F.~Saueressig, \emph{{Asymptotic safety in
  higher-derivative gravity}},
  \href{https://doi.org/10.1142/S0217732309031521}{\emph{Mod. Phys. Lett. A}
  {\bfseries 24} (2009) 2233}
  [\href{https://arxiv.org/abs/0901.2984}{{\ttfamily arXiv:0901.2984}}].

\bibitem{Eichhorn:2018yfc}
A.~Eichhorn, \emph{{An asymptotically safe guide to quantum gravity and
  matter}}, \href{https://doi.org/10.3389/fspas.2018.00047}{\emph{Front.
  Astron. Space Sci.} {\bfseries 5} (2019) 47}
  [\href{https://arxiv.org/abs/1810.07615}{{\ttfamily arXiv:1810.07615}}].

\bibitem{Donoghue:2019clr}
J.F.~Donoghue, \emph{{A Critique of the Asymptotic Safety Program}},
  \href{https://doi.org/10.3389/fphy.2020.00056}{\emph{Front. in Phys.}
  {\bfseries 8} (2020) 56} [\href{https://arxiv.org/abs/1911.02967}{{\ttfamily
  arXiv:1911.02967}}].

\bibitem{Stelle:1976gc}
K.S.~Stelle, \emph{{Renormalization of Higher Derivative Quantum Gravity}},
  \href{https://doi.org/10.1103/PhysRevD.16.953}{\emph{Phys. Rev. D} {\bfseries
  16} (1977) 953}.

\bibitem{Stelle:1977ry}
K.S.~Stelle, \emph{{Classical Gravity with Higher Derivatives}},
  \href{https://doi.org/10.1007/BF00760427}{\emph{Gen. Rel. Grav.} {\bfseries
  9} (1978) 353}.

\bibitem{Alvarez-Gaume:2015rwa}
L.~Alvarez-Gaume, A.~Kehagias, C.~Kounnas, D.~L\"ust and A.~Riotto,
  \emph{{Aspects of Quadratic Gravity}},
  \href{https://doi.org/10.1002/prop.201500100}{\emph{Fortsch. Phys.}
  {\bfseries 64} (2016) 176}
  [\href{https://arxiv.org/abs/1505.07657}{{\ttfamily arXiv:1505.07657}}].

\bibitem{Salvio:2018crh}
A.~Salvio, \emph{{Quadratic Gravity}},
  \href{https://doi.org/10.3389/fphy.2018.00077}{\emph{Front. in Phys.}
  {\bfseries 6} (2018) 77} [\href{https://arxiv.org/abs/1804.09944}{{\ttfamily
  arXiv:1804.09944}}].

\bibitem{Fradkin:1981iu}
E.S.~Fradkin and A.A.~Tseytlin, \emph{{Renormalizable asymptotically free
  quantum theory of gravity}},
  \href{https://doi.org/10.1016/0550-3213(82)90444-8}{\emph{Nucl. Phys. B}
  {\bfseries 201} (1982) 469}.

\bibitem{Avramidi:1985ki}
I.G.~Avramidi and A.O.~Barvinsky, \emph{{Asymptotic freedom in
  higher-derivative quantum gravity}},
  \href{https://doi.org/10.1016/0370-2693(85)90248-5}{\emph{Phys. Lett. B}
  {\bfseries 159} (1985) 269}.

\bibitem{Codello:2006in}
A.~Codello and R.~Percacci, \emph{{Fixed points of higher derivative gravity}},
  \href{https://doi.org/10.1103/PhysRevLett.97.221301}{\emph{Phys. Rev. Lett.}
  {\bfseries 97} (2006) 221301}
  [\href{https://arxiv.org/abs/hep-th/0607128}{{\ttfamily hep-th/0607128}}].

\bibitem{Niedermaier:2009zz}
M.R.~Niedermaier, \emph{{Gravitational Fixed Points from Perturbation Theory}},
  \href{https://doi.org/10.1103/PhysRevLett.103.101303}{\emph{Phys. Rev. Lett.}
  {\bfseries 103} (2009) 101303}.

\bibitem{Niedermaier:2010zz}
M.~Niedermaier, \emph{{Gravitational fixed points and asymptotic safety from
  perturbation theory}},
  \href{https://doi.org/10.1016/j.nuclphysb.2010.01.016}{\emph{Nucl. Phys. B}
  {\bfseries 833} (2010) 226}.

\bibitem{Ohta:2013uca}
N.~Ohta and R.~Percacci, \emph{{Higher Derivative Gravity and Asymptotic Safety
  in Diverse Dimensions}},
  \href{https://doi.org/10.1088/0264-9381/31/1/015024}{\emph{Class. Quant.
  Grav.} {\bfseries 31} (2014) 015024}
  [\href{https://arxiv.org/abs/1308.3398}{{\ttfamily arXiv:1308.3398}}].

\bibitem{Holdom:2015kbf}
B.~Holdom and J.~Ren, \emph{{QCD analogy for quantum gravity}},
  \href{https://doi.org/10.1103/PhysRevD.93.124030}{\emph{Phys. Rev. D}
  {\bfseries 93} (2016) 124030}
  [\href{https://arxiv.org/abs/1512.05305}{{\ttfamily arXiv:1512.05305}}].

\bibitem{Holdom:2016xfn}
B.~Holdom and J.~Ren, \emph{{Quadratic gravity: from weak to strong}},
  \href{https://doi.org/10.1142/S0218271816430045}{\emph{Int. J. Mod. Phys. D}
  {\bfseries 25} (2016) 1643004}
  [\href{https://arxiv.org/abs/1605.05006}{{\ttfamily arXiv:1605.05006}}].

\bibitem{Buccio:2024hys}
D.~Buccio, J.F.~Donoghue, G.~Menezes and R.~Percacci, \emph{{Physical Running
  of Couplings in Quadratic Gravity}},
  \href{https://doi.org/10.1103/PhysRevLett.133.021604}{\emph{Phys. Rev. Lett.}
  {\bfseries 133} (2024) 021604}
  [\href{https://arxiv.org/abs/2403.02397}{{\ttfamily arXiv:2403.02397}}].

\bibitem{Buccio:2024omv}
D.~Buccio, L.~Parente and O.~Zanusso, \emph{{Physical Running in Conformal
  Gravity and Higher Derivative Scalars}},
  \href{https://arxiv.org/abs/2410.21475}{{\ttfamily arXiv:2410.21475}}.

\bibitem{Kawai:2024aim}
H.~Kawai and N.~Ohta, \emph{{An observation on the beta functions in quadratic
  gravity}}, \href{https://doi.org/10.1016/j.physletb.2024.138863}{\emph{Phys.
  Lett. B} {\bfseries 855} (2024) 138863}
  [\href{https://arxiv.org/abs/2405.05706}{{\ttfamily arXiv:2405.05706}}].

\bibitem{Hinterbichler:2015soa}
K.~Hinterbichler and M.~Saravani, \emph{{St\"uckelberg approach to quadratic
  curvature gravity and its decoupling limits}},
  \href{https://doi.org/10.1103/PhysRevD.93.065006}{\emph{Phys. Rev. D}
  {\bfseries 93} (2016) 065006}
  [\href{https://arxiv.org/abs/1508.02401}{{\ttfamily arXiv:1508.02401}}].

\bibitem{Hawking:2001yt}
S.W.~Hawking and T.~Hertog, \emph{{Living with ghosts}},
  \href{https://doi.org/10.1103/PhysRevD.65.103515}{\emph{Phys. Rev. D}
  {\bfseries 65} (2002) 103515}
  [\href{https://arxiv.org/abs/hep-th/0107088}{{\ttfamily hep-th/0107088}}].

\bibitem{Mannheim:2004qz}
P.D.~Mannheim and A.~Davidson, \emph{{Dirac quantization of the Pais-Uhlenbeck
  fourth order oscillator}},
  \href{https://doi.org/10.1103/PhysRevA.71.042110}{\emph{Phys. Rev. A}
  {\bfseries 71} (2005) 042110}
  [\href{https://arxiv.org/abs/hep-th/0408104}{{\ttfamily hep-th/0408104}}].

\bibitem{Mannheim:2006rd}
P.D.~Mannheim, \emph{{Solution to the ghost problem in fourth order derivative
  theories}}, \href{https://doi.org/10.1007/s10701-007-9119-7}{\emph{Found.
  Phys.} {\bfseries 37} (2007) 532}
  [\href{https://arxiv.org/abs/hep-th/0608154}{{\ttfamily hep-th/0608154}}].

\bibitem{Bender:2007wu}
C.M.~Bender and P.D.~Mannheim, \emph{{No-ghost theorem for the fourth-order
  derivative Pais-Uhlenbeck oscillator model}},
  \href{https://doi.org/10.1103/PhysRevLett.100.110402}{\emph{Phys. Rev. Lett.}
  {\bfseries 100} (2008) 110402}
  [\href{https://arxiv.org/abs/0706.0207}{{\ttfamily arXiv:0706.0207}}].

\bibitem{Donoghue:2017fvm}
J.F.~Donoghue, \emph{{Quartic propagators, negative norms and the physical
  spectrum}}, \href{https://doi.org/10.1103/PhysRevD.96.044007}{\emph{Phys.
  Rev. D} {\bfseries 96} (2017) 044007}
  [\href{https://arxiv.org/abs/1704.01533}{{\ttfamily arXiv:1704.01533}}].

\bibitem{Donoghue:2018izj}
J.F.~Donoghue and G.~Menezes, \emph{{Gauge Assisted Quadratic Gravity: A
  Framework for UV Complete Quantum Gravity}},
  \href{https://doi.org/10.1103/PhysRevD.97.126005}{\emph{Phys. Rev. D}
  {\bfseries 97} (2018) 126005}
  [\href{https://arxiv.org/abs/1804.04980}{{\ttfamily arXiv:1804.04980}}].

\bibitem{Donoghue:2019fcb}
J.F.~Donoghue and G.~Menezes, \emph{{Unitarity, stability and loops of unstable
  ghosts}}, \href{https://doi.org/10.1103/PhysRevD.100.105006}{\emph{Phys. Rev.
  D} {\bfseries 100} (2019) 105006}
  [\href{https://arxiv.org/abs/1908.02416}{{\ttfamily arXiv:1908.02416}}].

\bibitem{Donoghue:2019ecz}
J.F.~Donoghue and G.~Menezes, \emph{{Arrow of Causality and Quantum Gravity}},
  \href{https://doi.org/10.1103/PhysRevLett.123.171601}{\emph{Phys. Rev. Lett.}
  {\bfseries 123} (2019) 171601}
  [\href{https://arxiv.org/abs/1908.04170}{{\ttfamily arXiv:1908.04170}}].

\bibitem{Donoghue:2020mdd}
J.F.~Donoghue and G.~Menezes, \emph{{Quantum causality and the arrows of time
  and thermodynamics}},
  \href{https://doi.org/10.1016/j.ppnp.2020.103812}{\emph{Prog. Part. Nucl.
  Phys.} {\bfseries 115} (2020) 103812}
  [\href{https://arxiv.org/abs/2003.09047}{{\ttfamily arXiv:2003.09047}}].

\bibitem{Donoghue:2021eto}
J.F.~Donoghue and G.~Menezes, \emph{{Ostrogradsky instability can be overcome
  by quantum physics}},
  \href{https://doi.org/10.1103/PhysRevD.104.045010}{\emph{Phys. Rev. D}
  {\bfseries 104} (2021) 045010}
  [\href{https://arxiv.org/abs/2105.00898}{{\ttfamily arXiv:2105.00898}}].

\bibitem{Donoghue:2021meq}
J.F.~Donoghue and G.~Menezes, \emph{{Causality and gravity}},
  \href{https://doi.org/10.1007/JHEP11(2021)010}{\emph{JHEP} {\bfseries 11}
  (2021) 010} [\href{https://arxiv.org/abs/2106.05912}{{\ttfamily
  arXiv:2106.05912}}].

\bibitem{Donoghue:2021cza}
J.F.~Donoghue and G.~Menezes, \emph{{On quadratic gravity}},
  \href{https://doi.org/10.1393/ncc/i2022-22026-7}{\emph{Nuovo Cim. C}
  {\bfseries 45} (2022) 26} [\href{https://arxiv.org/abs/2112.01974}{{\ttfamily
  arXiv:2112.01974}}].

\bibitem{Holdom:2023usn}
B.~Holdom, \emph{{Running couplings and unitarity in a 4-derivative scalar
  field theory}},
  \href{https://doi.org/10.1016/j.physletb.2023.138023}{\emph{Phys. Lett. B}
  {\bfseries 843} (2023) 138023}
  [\href{https://arxiv.org/abs/2303.06723}{{\ttfamily arXiv:2303.06723}}].

\bibitem{Holdom:2024cfq}
B.~Holdom, \emph{{UV-complete 4-derivative scalar field theory}},
  \href{https://doi.org/10.1016/j.nuclphysb.2024.116472}{\emph{Nucl. Phys. B}
  {\bfseries 1000} (2024) 116472}
  [\href{https://arxiv.org/abs/2402.09223}{{\ttfamily arXiv:2402.09223}}].

\bibitem{Holdom:2024onr}
B.~Holdom, \emph{{Making sense of ghosts}},
  \href{https://doi.org/10.1016/j.nuclphysb.2024.116696}{\emph{Nucl. Phys. B}
  {\bfseries 1008} (2024) 116696}
  [\href{https://arxiv.org/abs/2408.04089}{{\ttfamily arXiv:2408.04089}}].

\bibitem{Clunan:2009er}
T.~Clunan and M.~Sasaki, \emph{{Tensor ghosts in the inflationary cosmology}},
  \href{https://doi.org/10.1088/0264-9381/27/16/165014}{\emph{Class. Quant.
  Grav.} {\bfseries 27} (2010) 165014}
  [\href{https://arxiv.org/abs/0907.3868}{{\ttfamily arXiv:0907.3868}}].

\bibitem{Deruelle:2010kf}
N.~Deruelle, M.~Sasaki, Y.~Sendouda and A.~Youssef, \emph{{Inflation with a
  Weyl term, or ghosts at work}},
  \href{https://doi.org/10.1088/1475-7516/2011/03/040}{\emph{JCAP} {\bfseries
  03} (2011) 040} [\href{https://arxiv.org/abs/1012.5202}{{\ttfamily
  arXiv:1012.5202}}].

\bibitem{Deruelle:2012xv}
N.~Deruelle, M.~Sasaki, Y.~Sendouda and A.~Youssef, \emph{{Lorentz-violating vs
  ghost gravitons: the example of Weyl gravity}},
  \href{https://doi.org/10.1007/JHEP09(2012)009}{\emph{JHEP} {\bfseries 09}
  (2012) 009} [\href{https://arxiv.org/abs/1202.3131}{{\ttfamily
  arXiv:1202.3131}}].

\bibitem{Anselmi:2020lpp}
D.~Anselmi, E.~Bianchi and M.~Piva, \emph{{Predictions of quantum gravity in
  inflationary cosmology: effects of the Weyl-squared term}},
  \href{https://doi.org/10.1007/JHEP07(2020)211}{\emph{JHEP} {\bfseries 07}
  (2020) 211} [\href{https://arxiv.org/abs/2005.10293}{{\ttfamily
  arXiv:2005.10293}}].

\bibitem{Anselmi:2021dag}
D.~Anselmi, \emph{{Perturbation spectra and renormalization-group techniques in
  double-field inflation and quantum gravity cosmology}},
  \href{https://doi.org/10.1088/1475-7516/2021/07/037}{\emph{JCAP} {\bfseries
  07} (2021) 037} [\href{https://arxiv.org/abs/2105.05864}{{\ttfamily
  arXiv:2105.05864}}].

\bibitem{Anselmi:2021rye}
D.~Anselmi, F.~Fruzza and M.~Piva, \emph{{Renormalization-group techniques for
  single-field inflation in primordial cosmology and quantum gravity}},
  \href{https://doi.org/10.1088/1361-6382/ac2b07}{\emph{Class. Quant. Grav.}
  {\bfseries 38} (2021) 225011}
  [\href{https://arxiv.org/abs/2103.01653}{{\ttfamily arXiv:2103.01653}}].

\bibitem{DeFelice:2023psw}
A.~De~Felice, R.~Kawaguchi, K.~Mizui and S.~Tsujikawa, \emph{{Starobinsky
  inflation with a quadratic Weyl tensor}},
  \href{https://doi.org/10.1103/PhysRevD.108.123524}{\emph{Phys. Rev. D}
  {\bfseries 108} (2023) 123524}
  [\href{https://arxiv.org/abs/2309.01835}{{\ttfamily arXiv:2309.01835}}].

\bibitem{Salvio:2017xul}
A.~Salvio, \emph{{Inflationary Perturbations in No-Scale Theories}},
  \href{https://doi.org/10.1140/epjc/s10052-017-4825-6}{\emph{Eur. Phys. J. C}
  {\bfseries 77} (2017) 267}
  [\href{https://arxiv.org/abs/1703.08012}{{\ttfamily arXiv:1703.08012}}].

\bibitem{Salvio:2017qkx}
A.~Salvio and A.~Strumia, \emph{{Agravity up to infinite energy}},
  \href{https://doi.org/10.1140/epjc/s10052-018-5588-4}{\emph{Eur. Phys. J. C}
  {\bfseries 78} (2018) 124}
  [\href{https://arxiv.org/abs/1705.03896}{{\ttfamily arXiv:1705.03896}}].

\bibitem{Salvio:2019ewf}
A.~Salvio, \emph{{Metastability in Quadratic Gravity}},
  \href{https://doi.org/10.1103/PhysRevD.99.103507}{\emph{Phys. Rev. D}
  {\bfseries 99} (2019) 103507}
  [\href{https://arxiv.org/abs/1902.09557}{{\ttfamily arXiv:1902.09557}}].

\bibitem{Salvio:2024joi}
A.~Salvio, \emph{{A non-perturbative and background-independent formulation of
  quadratic gravity}},
  \href{https://doi.org/10.1088/1475-7516/2024/07/092}{\emph{JCAP} {\bfseries
  07} (2024) 092} [\href{https://arxiv.org/abs/2404.08034}{{\ttfamily
  arXiv:2404.08034}}].

\bibitem{Held:2023aap}
A.~Held and H.~Lim, \emph{{Nonlinear evolution of quadratic gravity in 3+1
  dimensions}}, \href{https://doi.org/10.1103/PhysRevD.108.104025}{\emph{Phys.
  Rev. D} {\bfseries 108} (2023) 104025}
  [\href{https://arxiv.org/abs/2306.04725}{{\ttfamily arXiv:2306.04725}}].

\bibitem{Lehners:2019ibe}
J.L.~Lehners and K.~Stelle, \emph{{A Safe Beginning for the Universe?}},
  \href{https://doi.org/10.1103/PhysRevD.100.083540}{\emph{Phys.\ Rev.\ D}
  {\bfseries 100} (2019) 083540}
  [\href{https://arxiv.org/abs/1909.01169}{{\ttfamily arXiv:1909.01169}}].

\bibitem{Lehners:2023fud}
J.L.~Lehners and K.S.~Stelle, \emph{{Higher-order gravity, finite action, and a
  safe beginning for the universe}},
  \href{https://doi.org/10.1140/epjp/s13360-024-05125-y}{\emph{Eur. Phys. J.
  Plus} {\bfseries 139} (2024) 380}
  [\href{https://arxiv.org/abs/2312.14048}{{\ttfamily arXiv:2312.14048}}].

\bibitem{Bonanno:2024fcv}
A.~Bonanno and S.~Silveravalle, \emph{{Black holes at a crossroads: late-stage
  evaporation in quadratic gravity}},  in \emph{{17th Marcel Grossmann
  Meeting}: {On Recent Developments in Theoretical and Experimental General
  Relativity, Gravitation, and Relativistic Field Theories}}, 2024,
  \href{https://arxiv.org/abs/2409.16690}{{\ttfamily arXiv:2409.16690}}.

\bibitem{Buoninfante:2024oyi}
L.~Buoninfante, F.~Di~Filippo, I.~Kol\'a\v{r} and F.~Saueressig, \emph{{Dust
  collapse and horizon formation in quadratic gravity}},
  \href{https://doi.org/10.1088/1475-7516/2025/01/114}{\emph{JCAP} {\bfseries
  01} (2025) 114} [\href{https://arxiv.org/abs/2410.05941}{{\ttfamily
  arXiv:2410.05941}}].

\bibitem{Belokurov:2021oer}
V.V.~Belokurov and E.T.~Shavgulidze, \emph{{Path integrals in quadratic
  gravity}}, \href{https://doi.org/10.1007/JHEP02(2022)112}{\emph{JHEP}
  {\bfseries 02} (2022) 112}
  [\href{https://arxiv.org/abs/2110.06041}{{\ttfamily arXiv:2110.06041}}].

\bibitem{Belokurov:2024pjr}
V.V.~Belokurov, V.V.~Chistiakov and E.T.~Shavgulidze, \emph{{Perturbation
  Theory for Path Integrals in Quadratic Gravity}},
  \href{https://arxiv.org/abs/2410.19457}{{\ttfamily arXiv:2410.19457}}.

\bibitem{Edelstein:2021jyu}
J.D.~Edelstein, R.~Ghosh, A.~Laddha and S.~Sarkar, \emph{{Causality constraints
  in Quadratic Gravity}},
  \href{https://doi.org/10.1007/JHEP09(2021)150}{\emph{JHEP} {\bfseries 09}
  (2021) 150} [\href{https://arxiv.org/abs/2107.07424}{{\ttfamily
  arXiv:2107.07424}}].

\bibitem{Edelstein:2024jzu}
J.D.~Edelstein, R.~Ghosh, A.~Laddha and S.~Sarkar, \emph{{Restoring Causality
  in Higher Curvature Gravity}},
  \href{https://arxiv.org/abs/2409.16935}{{\ttfamily arXiv:2409.16935}}.

\bibitem{Berglund:2022qcc}
P.~Berglund, L.~Freidel, T.~Hubsch, J.~Kowalski-Glikman, R.G.~Leigh,
  D.~Mattingly et~al., \emph{{Infrared Properties of Quantum Gravity: UV/IR
  Mixing, Gravitizing the Quantum -- Theory and Observation}},  in
  \emph{{Snowmass 2021}}, 2022,
  \href{https://arxiv.org/abs/2202.06890}{{\ttfamily arXiv:2202.06890}}.

\bibitem{Hawking:1976ra}
S.W.~Hawking, \emph{{Breakdown of Predictability in Gravitational Collapse}},
  \href{https://doi.org/10.1103/PhysRevD.14.2460}{\emph{Phys. Rev. D}
  {\bfseries 14} (1976) 2460}.

\bibitem{Mathur:2009hf}
S.D.~Mathur, \emph{{The Information paradox: A Pedagogical introduction}},
  \href{https://doi.org/10.1088/0264-9381/26/22/224001}{\emph{Class. Quant.
  Grav.} {\bfseries 26} (2009) 224001}
  [\href{https://arxiv.org/abs/0909.1038}{{\ttfamily arXiv:0909.1038}}].

\bibitem{Almheiri:2012rt}
A.~Almheiri, D.~Marolf, J.~Polchinski and J.~Sully, \emph{{Black Holes:
  Complementarity or Firewalls?}},
  \href{https://doi.org/10.1007/JHEP02(2013)062}{\emph{JHEP} {\bfseries 02}
  (2013) 062} [\href{https://arxiv.org/abs/1207.3123}{{\ttfamily
  arXiv:1207.3123}}].

\bibitem{Cardoso:2016rao}
V.~Cardoso, E.~Franzin and P.~Pani, \emph{{Is the gravitational-wave ringdown a
  probe of the event horizon?}},
  \href{https://doi.org/10.1103/PhysRevLett.116.171101}{\emph{Phys. Rev. Lett.}
  {\bfseries 116} (2016) 171101}
  [\href{https://arxiv.org/abs/1602.07309}{{\ttfamily arXiv:1602.07309}}].

\bibitem{Cardoso:2016oxy}
V.~Cardoso, S.~Hopper, C.F.B.~Macedo, C.~Palenzuela and P.~Pani,
  \emph{{Gravitational-wave signatures of exotic compact objects and of quantum
  corrections at the horizon scale}},
  \href{https://doi.org/10.1103/PhysRevD.94.084031}{\emph{Phys. Rev. D}
  {\bfseries 94} (2016) 084031}
  [\href{https://arxiv.org/abs/1608.08637}{{\ttfamily arXiv:1608.08637}}].

\bibitem{Abedi:2016hgu}
J.~Abedi, H.~Dykaar and N.~Afshordi, \emph{{Echoes from the Abyss: Tentative
  evidence for Planck-scale structure at black hole horizons}},
  \href{https://doi.org/10.1103/PhysRevD.96.082004}{\emph{Phys. Rev. D}
  {\bfseries 96} (2017) 082004}
  [\href{https://arxiv.org/abs/1612.00266}{{\ttfamily arXiv:1612.00266}}].

\bibitem{Cardoso:2017cqb}
V.~Cardoso and P.~Pani, \emph{{Tests for the existence of black holes through
  gravitational wave echoes}},
  \href{https://doi.org/10.1038/s41550-017-0225-y}{\emph{Nature Astron.}
  {\bfseries 1} (2017) 586} [\href{https://arxiv.org/abs/1709.01525}{{\ttfamily
  arXiv:1709.01525}}].

\bibitem{Wang:2018gin}
Q.~Wang and N.~Afshordi, \emph{{Black hole echology: The
  observer\textquoteright{}s manual}},
  \href{https://doi.org/10.1103/PhysRevD.97.124044}{\emph{Phys. Rev. D}
  {\bfseries 97} (2018) 124044}
  [\href{https://arxiv.org/abs/1803.02845}{{\ttfamily arXiv:1803.02845}}].

\bibitem{Abedi:2020ujo}
J.~Abedi, N.~Afshordi, N.~Oshita and Q.~Wang, \emph{{Quantum Black Holes in the
  Sky}}, \href{https://doi.org/10.3390/universe6030043}{\emph{Universe}
  {\bfseries 6} (2020) 43} [\href{https://arxiv.org/abs/2001.09553}{{\ttfamily
  arXiv:2001.09553}}].

\bibitem{Lu:2015cqa}
H.~Lu, A.~Perkins, C.N.~Pope and K.S.~Stelle, \emph{{Black Holes in
  Higher-Derivative Gravity}},
  \href{https://doi.org/10.1103/PhysRevLett.114.171601}{\emph{Phys. Rev. Lett.}
  {\bfseries 114} (2015) 171601}
  [\href{https://arxiv.org/abs/1502.01028}{{\ttfamily arXiv:1502.01028}}].

\bibitem{Lu:2015psa}
H.~L\"u, A.~Perkins, C.N.~Pope and K.S.~Stelle, \emph{{Spherically Symmetric
  Solutions in Higher-Derivative Gravity}},
  \href{https://doi.org/10.1103/PhysRevD.92.124019}{\emph{Phys. Rev. D}
  {\bfseries 92} (2015) 124019}
  [\href{https://arxiv.org/abs/1508.00010}{{\ttfamily arXiv:1508.00010}}].

\bibitem{Lu:2015tle}
H.~L\"u, A.~Perkins, C.N.~Pope and K.S.~Stelle, \emph{{Black holes in D = 4
  higher-derivative gravity}},
  \href{https://doi.org/10.1142/S0217751X15450165}{\emph{Int. J. Mod. Phys. A}
  {\bfseries 30} (2015) 1545016}.

\bibitem{Stelle:2017bdu}
K.S.~Stelle, \emph{{Abdus Salam and Quadratic Curvature Gravity: Classical
  Solutions}}, \href{https://doi.org/10.1142/S0217751X17410123}{\emph{Int. J.
  Mod. Phys. A} {\bfseries 32} (2017) 1741012}.

\bibitem{Lu:2017kzi}
H.~L\"u, A.~Perkins, C.N.~Pope and K.S.~Stelle, \emph{{Lichnerowicz Modes and
  Black Hole Families in Ricci Quadratic Gravity}},
  \href{https://doi.org/10.1103/PhysRevD.96.046006}{\emph{Phys. Rev. D}
  {\bfseries 96} (2017) 046006}
  [\href{https://arxiv.org/abs/1704.05493}{{\ttfamily arXiv:1704.05493}}].

\bibitem{Holdom:2016nek}
B.~Holdom and J.~Ren, \emph{{Not quite a black hole}},
  \href{https://doi.org/10.1103/PhysRevD.95.084034}{\emph{Phys. Rev. D}
  {\bfseries 95} (2017) 084034}
  [\href{https://arxiv.org/abs/1612.04889}{{\ttfamily arXiv:1612.04889}}].

\bibitem{Holdom:2022zzo}
B.~Holdom, \emph{{2-2-holes simplified}},
  \href{https://doi.org/10.1016/j.physletb.2022.137142}{\emph{Phys. Lett. B}
  {\bfseries 830} (2022) 137142}
  [\href{https://arxiv.org/abs/2202.08442}{{\ttfamily arXiv:2202.08442}}].

\bibitem{Holdom:2019bdv}
B.~Holdom, \emph{{Not quite black holes at LIGO}},
  \href{https://doi.org/10.1103/PhysRevD.101.064063}{\emph{Phys. Rev. D}
  {\bfseries 101} (2020) 064063}
  [\href{https://arxiv.org/abs/1909.11801}{{\ttfamily arXiv:1909.11801}}].

\bibitem{Aydemir:2020xfd}
U.~Aydemir, B.~Holdom and J.~Ren, \emph{{Not quite black holes as dark
  matter}}, \href{https://doi.org/10.1103/PhysRevD.102.024058}{\emph{Phys. Rev.
  D} {\bfseries 102} (2020) 024058}
  [\href{https://arxiv.org/abs/2003.10682}{{\ttfamily arXiv:2003.10682}}].

\bibitem{Holdom:2020onl}
B.~Holdom, \emph{{Damping of gravitational waves in 2-2-holes}},
  \href{https://doi.org/10.1016/j.physletb.2020.136023}{\emph{Phys. Lett. B}
  {\bfseries 813} (2021) 136023}
  [\href{https://arxiv.org/abs/2004.11285}{{\ttfamily arXiv:2004.11285}}].

\bibitem{Holdom:2022npq}
B.~Holdom, \emph{{UV completion and not quite black holes}},
  \href{https://doi.org/10.1393/ncc/i2022-22036-5}{\emph{Nuovo Cim. C}
  {\bfseries 45} (2022) 36}.

\bibitem{Holdom:2022fsm}
B.~Holdom, \emph{{Towards rotating 2-2-holes}},
  \href{https://arxiv.org/abs/2208.08461}{{\ttfamily arXiv:2208.08461}}.

\bibitem{Borissova:2022jqj}
J.N.~Borissova, A.~Held and N.~Afshordi, \emph{{Scale-invariance at the core of
  quantum black holes}},
  \href{https://doi.org/10.1088/1361-6382/acbc60}{\emph{Class. Quant. Grav.}
  {\bfseries 40} (2023) 075011}
  [\href{https://arxiv.org/abs/2203.02559}{{\ttfamily arXiv:2203.02559}}].

\bibitem{Kazanas:1988qa}
D.~Kazanas and P.D.~Mannheim, \emph{{General Structure of the Gravitational
  Equations of Motion in Conformal Weyl Gravity}},
  \href{https://doi.org/10.1086/191573}{\emph{Astrophys. J. Suppl.} {\bfseries
  76} (1991) 431}.

\bibitem{Mannheim:2011ds}
P.D.~Mannheim, \emph{{Making the Case for Conformal Gravity}},
  \href{https://doi.org/10.1007/s10701-011-9608-6}{\emph{Found. Phys.}
  {\bfseries 42} (2012) 388} [\href{https://arxiv.org/abs/1101.2186}{{\ttfamily
  arXiv:1101.2186}}].

\bibitem{Poisson:2009pwt}
E.~Poisson, \emph{{A Relativist's Toolkit: The Mathematics of Black-Hole
  Mechanics}}, Cambridge University Press (12, 2009),
  \href{https://doi.org/10.1017/CBO9780511606601}{10.1017/CBO9780511606601}.

\bibitem{Hawking:1984ph}
S.W.~Hawking and J.C.~Luttrell, \emph{{Higher Derivatives in Quantum Cosmology.
  1. The Isotropic Case}},
  \href{https://doi.org/10.1016/0550-3213(84)90380-8}{\emph{Nucl. Phys. B}
  {\bfseries 247} (1984) 250}.

\bibitem{Dyer:2008hb}
E.~Dyer and K.~Hinterbichler, \emph{{Boundary Terms, Variational Principles and
  Higher Derivative Modified Gravity}},
  \href{https://doi.org/10.1103/PhysRevD.79.024028}{\emph{Phys. Rev. D}
  {\bfseries 79} (2009) 024028}
  [\href{https://arxiv.org/abs/0809.4033}{{\ttfamily arXiv:0809.4033}}].

\bibitem{Deruelle:2009zk}
N.~Deruelle, M.~Sasaki, Y.~Sendouda and D.~Yamauchi, \emph{{Hamiltonian
  formulation of f(Riemann) theories of gravity}},
  \href{https://doi.org/10.1143/PTP.123.169}{\emph{Prog. Theor. Phys.}
  {\bfseries 123} (2010) 169}
  [\href{https://arxiv.org/abs/0908.0679}{{\ttfamily arXiv:0908.0679}}].

\bibitem{Hohm:2010jc}
O.~Hohm and E.~Tonni, \emph{{A boundary stress tensor for higher-derivative
  gravity in AdS and Lifshitz backgrounds}},
  \href{https://doi.org/10.1007/JHEP04(2010)093}{\emph{JHEP} {\bfseries 04}
  (2010) 093} [\href{https://arxiv.org/abs/1001.3598}{{\ttfamily
  arXiv:1001.3598}}].

\bibitem{Jonas:2021xkx}
C.~Jonas, J.L.~Lehners and J.~Quintin, \emph{{Cosmological consequences of a
  principle of finite amplitudes}},
  \href{https://doi.org/10.1103/PhysRevD.103.103525}{\emph{Phys. Rev. D}
  {\bfseries 103} (2021) 103525}
  [\href{https://arxiv.org/abs/2102.05550}{{\ttfamily arXiv:2102.05550}}].

\bibitem{Barrow:2019gzc}
J.D.~Barrow, \emph{{Finite Action Principle Revisited}},
  \href{https://doi.org/10.1103/PhysRevD.101.023527}{\emph{Phys. Rev. D}
  {\bfseries 101} (2020) 023527}
  [\href{https://arxiv.org/abs/1912.12926}{{\ttfamily arXiv:1912.12926}}].

\bibitem{Israel:1966rt}
W.~Israel, \emph{{Singular hypersurfaces and thin shells in general
  relativity}}, \href{https://doi.org/10.1007/BF02710419}{\emph{Nuovo Cim. B}
  {\bfseries 44S10} (1966) 1}.

\bibitem{Reina:2015gxa}
B.~Reina, J.M.M.~Senovilla and R.~Vera, \emph{{Junction conditions in quadratic
  gravity: thin shells and double layers}},
  \href{https://doi.org/10.1088/0264-9381/33/10/105008}{\emph{Class. Quant.
  Grav.} {\bfseries 33} (2016) 105008}
  [\href{https://arxiv.org/abs/1510.05515}{{\ttfamily arXiv:1510.05515}}].

\bibitem{Chu:2021uec}
C.S.~Chu and H.S.~Tan, \emph{{Generalized Darmois\textendash{}Israel Junction
  Conditions}}, \href{https://doi.org/10.3390/universe8050250}{\emph{Universe}
  {\bfseries 8} (2022) 250} [\href{https://arxiv.org/abs/2103.06314}{{\ttfamily
  arXiv:2103.06314}}].

\bibitem{Lehners:2023yrj}
J.L.~Lehners, \emph{{Review of the no-boundary wave function}},
  \href{https://doi.org/10.1016/j.physrep.2023.06.002}{\emph{Phys. Rept.}
  {\bfseries 1022} (2023) 1}
  [\href{https://arxiv.org/abs/2303.08802}{{\ttfamily arXiv:2303.08802}}].

\bibitem{Hartle:1983ai}
J.B.~Hartle and S.W.~Hawking, \emph{{Wave Function of the Universe}},
  \href{https://doi.org/10.1103/PhysRevD.28.2960}{\emph{Phys. Rev. D}
  {\bfseries 28} (1983) 2960}.

\bibitem{Vilenkin:1982de}
A.~Vilenkin, \emph{{Creation of Universes from Nothing}},
  \href{https://doi.org/10.1016/0370-2693(82)90866-8}{\emph{Phys. Lett. B}
  {\bfseries 117} (1982) 25}.

\bibitem{Vilenkin:1983xq}
A.~Vilenkin, \emph{{The Birth of Inflationary Universes}},
  \href{https://doi.org/10.1103/PhysRevD.27.2848}{\emph{Phys. Rev. D}
  {\bfseries 27} (1983) 2848}.

\bibitem{Vilenkin:1984wp}
A.~Vilenkin, \emph{{Quantum Creation of Universes}},
  \href{https://doi.org/10.1103/PhysRevD.30.509}{\emph{Phys. Rev. D} {\bfseries
  30} (1984) 509}.

\bibitem{Linde:1983mx}
A.D.~Linde, \emph{{Quantum Creation of the Inflationary Universe}},
  \href{https://doi.org/10.1007/BF02790571}{\emph{Lett. Nuovo Cim.} {\bfseries
  39} (1984) 401}.

\bibitem{Feldbrugge:2017fcc}
J.~Feldbrugge, J.L.~Lehners and N.~Turok, \emph{{No smooth beginning for
  spacetime}},
  \href{https://doi.org/10.1103/PhysRevLett.119.171301}{\emph{Phys. Rev. Lett.}
  {\bfseries 119} (2017) 171301}
  [\href{https://arxiv.org/abs/1705.00192}{{\ttfamily arXiv:1705.00192}}].

\bibitem{Gibbons:1976ue}
G.W.~Gibbons and S.W.~Hawking, \emph{{Action Integrals and Partition Functions
  in Quantum Gravity}},
  \href{https://doi.org/10.1103/PhysRevD.15.2752}{\emph{Phys. Rev. D}
  {\bfseries 15} (1977) 2752}.

\bibitem{Mathur:2024mvo}
S.D.~Mathur and M.~Mehta, \emph{{The universal thermodynamic properties of
  extremely compact objects}},
  \href{https://doi.org/10.1088/1361-6382/ad869e}{\emph{Class. Quant. Grav.}
  {\bfseries 41} (2024) 235011}
  [\href{https://arxiv.org/abs/2402.13166}{{\ttfamily arXiv:2402.13166}}].

\bibitem{Kontsevich:2021dmb}
M.~Kontsevich and G.~Segal, \emph{{Wick Rotation and the Positivity of Energy
  in Quantum Field Theory}},
  \href{https://doi.org/10.1093/qmath/haab027}{\emph{Quart. J. Math. Oxford
  Ser.} {\bfseries 72} (2021) 673}
  [\href{https://arxiv.org/abs/2105.10161}{{\ttfamily arXiv:2105.10161}}].

\bibitem{Louko:1995jw}
J.~Louko and R.D.~Sorkin, \emph{{Complex actions in two-dimensional topology
  change}}, \href{https://doi.org/10.1088/0264-9381/14/1/018}{\emph{Class.
  Quant. Grav.} {\bfseries 14} (1997) 179}
  [\href{https://arxiv.org/abs/gr-qc/9511023}{{\ttfamily gr-qc/9511023}}].

\bibitem{Witten:2021nzp}
E.~Witten, \emph{A note on complex spacetime metrics},  in \emph{Frank Wilczek:
  50 years of theoretical physics}, 2022,
  \href{https://arxiv.org/abs/2111.06514}{{\ttfamily arXiv:2111.06514}}.

\bibitem{Lehners:2021mah}
J.L.~Lehners, \emph{{Allowable complex metrics in minisuperspace quantum
  cosmology}}, \href{https://doi.org/10.1103/PhysRevD.105.026022}{\emph{Phys.
  Rev. D} {\bfseries 105} (2022) 026022}
  [\href{https://arxiv.org/abs/2111.07816}{{\ttfamily arXiv:2111.07816}}].

\bibitem{Jonas:2022uqb}
C.~Jonas, J.L.~Lehners and J.~Quintin, \emph{{Uses of complex metrics in
  cosmology}}, \href{https://doi.org/10.1007/JHEP08(2022)284}{\emph{JHEP}
  {\bfseries 08} (2022) 284}
  [\href{https://arxiv.org/abs/2205.15332}{{\ttfamily arXiv:2205.15332}}].

\bibitem{Lehners:2022xds}
J.L.~Lehners, \emph{{Allowable complex scalars from Kaluza-Klein
  compactifications and metric rescalings}},
  \href{https://doi.org/10.1103/PhysRevD.107.046004}{\emph{Phys. Rev. D}
  {\bfseries 107} (2023) 046004}
  [\href{https://arxiv.org/abs/2209.14669}{{\ttfamily arXiv:2209.14669}}].

\bibitem{Hertog:2023vot}
T.~Hertog, O.~Janssen and J.~Karlsson, \emph{{Kontsevich-Segal Criterion in the
  No-Boundary State Constrains Inflation}},
  \href{https://doi.org/10.1103/PhysRevLett.131.191501}{\emph{Phys. Rev. Lett.}
  {\bfseries 131} (2023) 191501}
  [\href{https://arxiv.org/abs/2305.15440}{{\ttfamily arXiv:2305.15440}}].

\bibitem{Lehners:2023pcn}
J.L.~Lehners and J.~Quintin, \emph{{A small Universe}},
  \href{https://doi.org/10.1016/j.physletb.2024.138488}{\emph{Phys. Lett. B}
  {\bfseries 850} (2024) 138488}
  [\href{https://arxiv.org/abs/2309.03272}{{\ttfamily arXiv:2309.03272}}].

\bibitem{Maldacena:2024uhs}
J.~Maldacena, \emph{{Comments on the no boundary wavefunction and slow roll
  inflation}},  \href{https://arxiv.org/abs/2403.10510}{{\ttfamily
  arXiv:2403.10510}}.

\bibitem{Janssen:2024vjn}
O.~Janssen, \emph{{KSW criterion in large field models}},
  \href{https://doi.org/10.1088/1361-6382/ad805d}{\emph{Class. Quant. Grav.}
  {\bfseries 41} (2024) 227001}
  [\href{https://arxiv.org/abs/2406.08422}{{\ttfamily arXiv:2406.08422}}].

\bibitem{Chakravarty:2024bna}
J.~Chakravarty, A.~Maloney, K.~Namjou and S.F.~Ross, \emph{{A new observable
  for holographic cosmology}},
  \href{https://doi.org/10.1007/JHEP10(2024)184}{\emph{JHEP} {\bfseries 10}
  (2024) 184} [\href{https://arxiv.org/abs/2407.04781}{{\ttfamily
  arXiv:2407.04781}}].

\bibitem{Hertog:2024nbh}
T.~Hertog, O.~Janssen and J.~Karlsson, \emph{{Kontsevich-Segal criterion in the
  no-boundary state constrains anisotropy}},
  \href{https://doi.org/10.1103/PhysRevD.111.046008}{\emph{Phys. Rev. D}
  {\bfseries 111} (2025) 046008}
  [\href{https://arxiv.org/abs/2408.02652}{{\ttfamily arXiv:2408.02652}}].

\bibitem{Feldbrugge:2017kzv}
J.~Feldbrugge, J.L.~Lehners and N.~Turok, \emph{{Lorentzian Quantum
  Cosmology}}, \href{https://doi.org/10.1103/PhysRevD.95.103508}{\emph{Phys.
  Rev. D} {\bfseries 95} (2017) 103508}
  [\href{https://arxiv.org/abs/1703.02076}{{\ttfamily arXiv:1703.02076}}].

\bibitem{Chen:2023hra}
Y.~Chen, V.~Ivo and J.~Maldacena, \emph{{Comments on the double cone
  wormhole}}, \href{https://doi.org/10.1007/JHEP04(2024)124}{\emph{JHEP}
  {\bfseries 04} (2024) 124}
  [\href{https://arxiv.org/abs/2310.11617}{{\ttfamily arXiv:2310.11617}}].

\bibitem{Bars:2013yba}
I.~Bars, P.~Steinhardt and N.~Turok, \emph{{Local Conformal Symmetry in Physics
  and Cosmology}},
  \href{https://doi.org/10.1103/PhysRevD.89.043515}{\emph{Phys. Rev. D}
  {\bfseries 89} (2014) 043515}
  [\href{https://arxiv.org/abs/1307.1848}{{\ttfamily arXiv:1307.1848}}].

\bibitem{Bars:2013vba}
I.~Bars, P.J.~Steinhardt and N.~Turok, \emph{{Cyclic Cosmology, Conformal
  Symmetry and the Metastability of the Higgs}},
  \href{https://doi.org/10.1016/j.physletb.2013.08.071}{\emph{Phys. Lett. B}
  {\bfseries 726} (2013) 50} [\href{https://arxiv.org/abs/1307.8106}{{\ttfamily
  arXiv:1307.8106}}].

\bibitem{Boyle:2018tzc}
L.~Boyle, K.~Finn and N.~Turok, \emph{{CPT-Symmetric Universe}},
  \href{https://doi.org/10.1103/PhysRevLett.121.251301}{\emph{Phys. Rev. Lett.}
  {\bfseries 121} (2018) 251301}
  [\href{https://arxiv.org/abs/1803.08928}{{\ttfamily arXiv:1803.08928}}].

\bibitem{Boyle:2021jej}
L.~Boyle and N.~Turok, \emph{{Two-Sheeted Universe, Analyticity and the Arrow
  of Time}},  \href{https://arxiv.org/abs/2109.06204}{{\ttfamily
  arXiv:2109.06204}}.

\bibitem{Boyle:2021jaz}
L.~Boyle and N.~Turok, \emph{{Cancelling the vacuum energy and Weyl anomaly in
  the standard model with dimension-zero scalar fields}},
  \href{https://arxiv.org/abs/2110.06258}{{\ttfamily arXiv:2110.06258}}.

\bibitem{Turok:2022fgq}
N.~Turok and L.~Boyle, \emph{{Gravitational entropy and the flatness,
  homogeneity and isotropy puzzles}},
  \href{https://doi.org/10.1016/j.physletb.2024.138443}{\emph{Phys. Lett. B}
  {\bfseries 849} (2024) 138443}
  [\href{https://arxiv.org/abs/2201.07279}{{\ttfamily arXiv:2201.07279}}].

\bibitem{Boyle:2022lcq}
L.~Boyle and N.~Turok, \emph{{Thermodynamic solution of the homogeneity,
  isotropy and flatness puzzles (and a clue to the cosmological constant)}},
  \href{https://doi.org/10.1016/j.physletb.2024.138442}{\emph{Phys. Lett. B}
  {\bfseries 849} (2024) 138442}
  [\href{https://arxiv.org/abs/2210.01142}{{\ttfamily arXiv:2210.01142}}].

\bibitem{Turok:2023amx}
N.~Turok and L.~Boyle, \emph{{A Minimal Explanation of the Primordial
  Cosmological Perturbations}},
  \href{https://arxiv.org/abs/2302.00344}{{\ttfamily arXiv:2302.00344}}.

\bibitem{Marolf:2020xie}
D.~Marolf and H.~Maxfield, \emph{{Transcending the ensemble: baby universes,
  spacetime wormholes, and the order and disorder of black hole information}},
  \href{https://doi.org/10.1007/JHEP08(2020)044}{\emph{JHEP} {\bfseries 08}
  (2020) 044} [\href{https://arxiv.org/abs/2002.08950}{{\ttfamily
  arXiv:2002.08950}}].

\bibitem{VanRaamsdonk:2020tlr}
M.~Van~Raamsdonk, \emph{{Comments on wormholes, ensembles, and cosmology}},
  \href{https://doi.org/10.1007/JHEP12(2021)156}{\emph{JHEP} {\bfseries 12}
  (2021) 156} [\href{https://arxiv.org/abs/2008.02259}{{\ttfamily
  arXiv:2008.02259}}].

\end{thebibliography}\endgroup

\end{document}